# Evaluation of a meta-analysis of ambient air quality as a risk factor for asthma exacerbation

Warren B. Kindzierski,[a*] S. Stanley Young,[b] Terry G. Meyer[c] and John D. Dunn[d]

[a]School of Public Health, University of Alberta, Edmonton, Alberta, Canada

[b]CGStat, Raleigh, NC, USA

[c]Outcome Based Medicine, Raleigh, NC, USA

[d]401 Rocky Hill Road, Brownwood, TX, USA

*Corresponding author: School of Public Health, University of Alberta, 3-57 South Academic Building, 11405-87 Avenue, Edmonton, Alberta, T6G 1C9 Canada; warrenk@ualberta.ca

# Abstract

False-positive results and bias may be common features of the biomedical literature today, including risk factor−chronic disease research. A study was undertaken to assess the reliability of base studies used in a meta-analysis examining whether carbon monoxide, particulate matter 10 μm and 2.5 μm (PM10 and PM2.5), sulfur dioxide, nitrogen dioxide and ozone are risk factors for asthma exacerbation (hospital admission and emergency room visits for asthma attack). The number of statistical tests and models were counted in 17 randomly selected base papers from 87 used in the meta-analysis. P-value plots for each air component were constructed to evaluate the effect heterogeneity of p-values used from all 87 base papers The number of statistical tests possible in the 17 selected base papers was large, median = 15,360 (interquartile range = 1,536 – 40,960), in comparison to results presented. Each p-value plot showed a two-component mixture with small p-values <.001 while other p-values appeared random (p-values >.05). Given potentially large numbers of statistical tests conducted in the 17 selected base papers, p-hacking cannot be ruled out as explanations for small p-values. Our interpretation of the meta-analysis is that the random p-values indicating null associations are more plausible and that the meta-analysis will not likely replicate in the absence of bias. We conclude the meta-analysis and base papers used are unreliable and do not offer evidence of value to inform public health practitioners about air quality as a risk factor for asthma exacerbation. We see the following areas crucial for enabling improvements in risk factor−chronic disease observational studies at the funding agency and journal level: preregistration, changes in funding agency and journal editor (and reviewer) practices, open sharing of data and facilitation of reproducibility research.

**Introduction**

*False-positives and bias in biomedical literature*

*False positives* – False-positive results may be a common feature of biomedical literature with others estimating that a majority of positive (statistically significant) research findings that physicians rely on – as much as 90 percent – may be flawed (Freedman 2010) and nothing more than 'false positives' (Keown 2012). Today the traditional .05 p-value has lost its ability to discriminate important biomedical research findings, especially when numerous comparisons (multiple tests) are being made, and can produce research results that are irreproducible and in some examples harmful to the public's health (Bock 2016). In addition, despite more than five decades of research and thousands of published epidemiologic studies on a wide array of chemicals in the environment, LaKind et al. (2015) indicate that it is often impossible to draw robust conclusions about the presence or absence of causal links between exposure – real or inferred – to specific environmental risk factors and human health effects.

False-positive findings can arise in research when statistical methods are applied incorrectly or when p-values are interpreted without sufficient understanding of the multiple testing problem (Forstmeier et al. 2017). False-positive results are reported by others to dominate the epidemiologic literature, with Ioannidis et al. (2011) suggesting that most of the traditional areas of epidemiologic research more closely reflect the performance settings and practices of early human genome epidemiology… showing at least 20 false-positive results for every one true-positive result. On the other hand, studies with negative (non-statistically significant) results are more likely to remain unpublished than studies with positive results (Franco et al. 2014). Epidemiological literature may be distorted as a result and any systematic review or meta-analysis of these studies would be biased (Egger et al. 2001, Sterne et al. 2001) because they are summarizing information and data from a misleading, selected body of evidence (Ioannidis 2008a, NASEM 2019).

Observational studies, where investigators observe individuals without manipulation or intervention, are a big part of the epidemiologic literature. However, far too many of the results coming from observational studies turn out to be wrong when they are carefully re-examined (Young and Karr 2011). This excess of false-positive results in published observational (and experimental) literature has been attributed mostly to bias (Taubes 1995, Boffetta et al. 2008, Ioannidis 2011a, Tsilidis et al. 2013, Bruns and Ioannidis 2016).

*Bias* – Bias consists of systematic alteration of research findings due to factors related to study design, data acquisition, analysis or reporting of results (Boffetta et al. 2008). Gotzsche (2006) suspects that selective reporting proliferates in published observational studies where researchers routinely test many models and questions during a study and then report those models that offer interesting (statistically significant) results. Further, selective reporting biases affecting overall results and specific analyses within studies are reported by Chan et al. (2004a,b), Chan and Altman (2005), Mathieu et al. (2009) and Ioannidis (2010) to likely be the greatest and most elusive issue distorting published research findings in the biomedical field today.

There are at least 235 catalogued biases in published biomedical literature (Chavalarias and Ioannidis 2010). Although most of the listed terminology for bias in literature is used infrequently, different bias terms used in diverse disciplines appear to refer to biases that have overlapping/similar meaning. Some of the very common bias terms used in current biomedical literature include (Chavalarias and Ioannidis 2010):

- Publication bias (tendency of authors, journal reviewers and editors to only report and publish positive or statistically significant results over null or non-statistically significant results). For example, Turner et al. (2008) reported that 31% of a cohort of studies for antidepressant drugs registered and reported to the Food and Drug Administration were never published. Whereas the published literature included 91% positive studies, the larger FDA cohort only contained 51% positive studies.
- Confounding (due either to incomplete statistical adjustment for measured variables or from the inability to adjust for unmeasured distorting variables). For example, the case cross-over study design attempts to capture an outcome of interest for each study participant in a observational study – e.g., occurrence of a short-term health episode such as an asthma exacerbation – when exposed and when unexposed (Howards 2018). Fixed risk factors for the outcome, such as genetic factors, do not change over time and therefore are the same when a participant is exposed and when unexposed (i.e., they do not confound the results). However, confounding can occur if there are risk factors that change over time (e.g., exposure to aeroallergens and viruses, changes in weather, etc.).

- Selection bias (use of improper procedures for selecting a sample population or as a result of factors that influence participation of the subjects in a study). For example, one form of selection bias commonly seen in cardiovascular studies is referred to as 'recurrent event bias' (Labos and Thanassoulis 2018), which occurs when a population under study is not representative of the general population but is selected to include patients who already have, in this case, cardiovascular disease.
- Response bias (conditions or factors that occur during the process of subjects responding to surveys affecting the way responses are provided). For example, people with a disease might be more likely to recall possible exposures to hazards in a health survey; or motivation of people to voluntarily participate in surveys about exposures to hazards is likely to be related to their prior beliefs and/or awareness of possibly relevant exposures (Joffe 2003).

*Positive and negative predictive values of risk factor−chronic disease effects*

Because of the prominence of disease prevention in our current health care system, observational risk factor−chronic disease research plays a key role in providing evidence to public health decision makers. Analysis of biomedical diagnostic test results being true depends on sensitivity, specificity and disease prevalence in a population (Schechter 1986, Last 2001, Shah 2003). Ioannidis (2005) incorporated the role of bias into this analysis and further developed relationships for understanding the probability of a research finding in a study being true for traditional epidemiological studies in the presence of bias. Given the importance of bias in observational studies (Ioannidis 2008a, 2011b), we extend the work of Ioannidis (2005) to illustrate the probability of positive research findings in a study (Positive Predictive Value or PPV) and negative, null, research findings in a study (Negative Predictive Value or NPV) being true for different levels of bias and prevalence rates for common chronic diseases in United States.

Table 1 presents estimates of prevalence for common chronic diseases of interest in United States (Supplemental Information (SI) 1 provides details upon which estimates are based). The four most common types of chronic diseases in the world's population (including developed and developing countries) are (Beaglehole et al. 2007, Alwan and MacLean 2009): respiratory diseases (primarily asthma), heart and stroke disease, diabetes and cancers. These diseases share key modifiable and preventable risk factors related to individual behavior (unhealthy diet, physical inactivity, tobacco use and harmful substance abuse, e.g., alcohol).

Using relationships we develop in SI 1, Figure 1 illustrates the probability of a research finding in a study being true as a function of disease prevalence within the range 0.0001–0.1 for two levels of bias (0.2−a study influenced by relatively minor bias and 0.8−a study influenced by relatively major bias). Bias, defined here after Ioannidis (2005), represents the *proportion of probed analyses* [relationships] *in a study that would not have been "research findings," but nevertheless end up presented and reported as such, because of bias*. All of the common diseases listed in Table 1 correspond with low post-study probabilities of a positive research finding in a study being true – less than 30%. On the other hand, these diseases correspond with very high post-study probabilities of a negative (null) research finding in a study being true – greater than 95%.

Figure 1 implies that a null risk factor−chronic disease finding in an observational study is very likely true; whereas, a positive risk factor−chronic disease finding is more likely to be false. For the most common chronic diseases of interest – i.e., diseases with a prevalence <0.1 – NPV (PPV) is essentially independent (dependent) of prevalence and bias. Also in Figure 1, a PPV exceeding 30% is difficult to achieve in risk factor−chronic disease epidemiological research for the most common chronic diseases of interest. A majority of modern biomedical research making claims is operating in areas with very low pre- and post-study probability for true findings (Ioannidis 2005). This is particularly true for the chronic diseases listed in Table 1 and, in part, it helps explain why false-positive results likely dominate the epidemiologic literature in this area. This is consistent with Ioannidis et al. (2011) reporting that most traditional fields of epidemiologic research have high ratios of false-positive to false-negative findings.

*Ambient air quality−chronic disease observational studies*

Many published observational epidemiology studies claim that certain air components are risk factors that are causal of chronic diseases (e.g., Chen and Goldberg 2009, To et al. 2015a,b). Claims that ambient air quality causes all of the chronic diseases listed in Table 1 have been made in the past, using meta-analysis – e.g., asthma (Zheng et al. 2015), COPD (DeVries et al. 2017), heart attack (Mustafic et al. 2012), diabetes (Eze et al. 2015), and breast (Keramatinia et al. 2016), prostate (Parent et al. 2013), colorectal (Turner et al. 2017) and lung cancer (Hamra et al. 2015). However it has been noted elsewhere (Cox 2017) that, in the presence of unmeasured confounders, causality is hard to establish for a relatively weak health risk factor such as ambient air quality. Low post-study probabilities for the common chronic diseases (see Table 1 and Figure 1) offer statistical reasons to question the reliability of these claims.

Part of this 'reliability" problem is that researchers of modern ambient air quality−effect observational studies can perform large numbers of statistical tests and use multiple statistical models – referred to here as MTMM (multiple testing and multiple modelling) (Westfall and Young 1993, Clyde 2000, Young 2017, Young and Kindzierski 2019):

- *Multiple testing* involves statistical null hypothesis testing of many separate predictor (e.g., air quality) variables against numerous representations of dependent (e.g., chronic disease) variables taking into account covariates, which may or may not act as confounders. For example, different air quality predictor variables − nitrogen dioxide, carbon monoxide − can be tested in the presence/absence of weather variables (e.g., temperature, relative humidity, etc.) against effects (e.g., heart attack hospitalizations) in a whole population of interest, females only, males only, those greater than 55 years old, etc.
- *Multiple modelling* involves statistical testing using multiple model selection procedures or different model forms (e.g., simple univariate, bivariate or multivariate logistic regression, times series, case crossover, etc.). For example, different models can be used in a single study to test independent predictor variables and covariates against dependent (e.g., chronic disease) variables.

MTMM is analogous to what Koop and Tole (2004) and Koop et al. (2010) call 'Model Selection' in statistical models used to estimate the effects of environmental impacts, referring to the fact that with $k$ potential explanatory variables in a model one can in principle run $2^k$ possible regressions and select one after the fact that that yields the 'best' result.

p-Hacking is multiple testing and multiple modelling without any statistical correction (Ellenberg 2014, Hubbard 2015, Chambers 2017, Harris 2017). It involves the relentless search for statistical significance and comes in many forms (Streiner 2018). p-Hacking enables researchers to find statistically significant results even when their samples are much too small to reliably detect the effect they are studying or even when they are studying an effect that is non-existent (Simonsohn et al. 2014). Motulsky (2014) offers some examples of different forms of p-hacking that can be used during a study: increasing sample size, analysing data subsets, increasing variables in a model, adjusting data, transforming data (i.e., log transformations), removing suspicious outliers, changing the control group, using different statistical tests.

***Objective of the current study***

Epidemiological study of chronic diseases with low prevalence operates in areas with very low pre- and post-study probability of positive research findings being true (Figure 1). Young and Kindzierski (2019) previously reported on the potential for epidemiology literature, particularly observational studies related to air quality component−heart attack associations, to be compromised by false positives and bias. For the present study we were interested in exploring whether the same sort of issues might be occurring with observational studies of air quality component−asthma attack associations. Asthma has an estimated prevalence in United States of 0.079 (Table 1) with any ambient air quality−chronic disease observational study 'positive effect' having a low post-study probability of being true – less than 25% (Figure 1).

     A meta-analysis offers a window into a research claim, for example, that some ambient air quality components, e.g., airborne fine particulate matter, are causal of a chronic disease. A meta-analysis examines a claim by taking a summary statistic along with a measure of its reliability from multiple individual ambient air quality−chronic disease studies (base papers) found in the epidemiological literature. These statistics are combined to give what is supposed to be a more reliable estimate of an air quality effect. A key assumption of a meta-analysis is that estimates drawn from the base papers for the analysis are unbiased estimates of the effect of interest (Boos and Stefanski 2013). However, as stated previously, studies with negative results are more likely to remain unpublished than studies with positive results leading to distortion of effects in the epidemiological literature and subsequent unreliable meta-analysis of these effects (Ioannidis 2008a).

     Here we evaluated the meta-analysis study of Zheng et al. (2015) [180 Google Scholar citations as of December 16, 2019]. We were keen in evaluating two properties

of this meta-analysis following along the lines of two recent published studies (Young et al. 2019 and Young and Kindzierski 2019):

- Whether claims in the base papers of the Zheng et al. meta-analysis are unreliable due to the presence of multiple testing and multiple modelling (MTMM) which can give rise to false positive results (Westfall and Young 1993, Young 2017).
- Whether heterogeneity across the base papers of the Zheng et al. meta-analysis is more complex than simple sampling from a single normal process (DerSimonian and Laird 1986).

**Methods**

Zheng et al. undertook a systematic computerized search of published observational studies to identify those studies focusing on short-term exposures (same day and lags up to 7 days; which are never chosen *a priori*) to six ambient air quality components (carbon monoxide (CO), particulate matter with aerodynamic equivalent diameter ≤10 micron (PM10), particulate matter with aerodynamic equivalent diameter ≤2.5 micron (PM2.5), sulfur dioxide (SO2), nitrogen dioxide (NO2) and ozone (O3)) and asthma exacerbation (hospital admission and emergency room visits for asthma attack).

Associations between air quality components and asthma-related hospital admission and emergency room visits were expressed as risk ratios (RRs) and 95% confidence intervals (CIs) that were derived from single-pollutant models reporting RRs (95% CIs) or percentage change (95% CIs). They further recalculated these associations to represent a 10 μg/m$^3$ increase, except for CO (where they recalculated associations to represent a 1 mg/m$^3$ increase).

Zhang et al. concluded that air all six air quality components were associated with significantly increased risks of asthma-related hospital admission and emergency room visits for all air quality components:

- CO: RR=1.045 (95% CI 1.029, 1.061)
- PM10: RR=1.010 (95% CI 1.008, 1.013)
- PM2.5: RR=1.023 (95% CI 1.015, 1.031)
- SO2: RR=1.011 (95% CI 95% CI 1.007, 1.015)
- NO2: RR=1.018 (95% CI 1.014, 1.022)

- O3: RR=1.009 (95% CI 1.006, 1.011)

It is generally accepted that there are two classes of causes of asthma – primary and secondary (Pekkanen and Pearce 1999). 'Primary' causes of asthma relate to the increase in risk of developing the disorder (e.g., asthma). Whereas 'secondary' causes relate to triggering/precipitation of asthma exacerbation. The Zheng et al. meta-analysis focused studies of short-term exposures to ambient air quality components and hence on secondary causes of asthma exacerbation.

Zheng et al. initially identified 1,099 literature reports. After screening titles and abstracts of these reports, they selected and assessed 246 full-text articles for eligibility, of which 87 base papers were selected for their meta-analysis. Citations and summary details for the 87 base papers each used by Zheng et al. in their meta-analysis are provided in SI 2.

*Analysis search space*

Analysis search space (or search space counts) represents an estimate of the number of statistical tests, in this case, of exposure-disease combinations tested in an observational study. Our interest in search space counts is explained further. During a study there is flexibility available to researchers to undertake a range of statistical tests and use different statistical models (i.e., perform MTMM) before selecting, using and reporting only a portion of the test and model results. This researcher flexibility is commonly referred to as *researcher degrees of freedom* in the psychological sciences (Wicherts et al. 2016).

Base papers with large search space counts suggest the use of a large number of statistical tests and statistical models and the potential for researchers to search through and only report only a portion of their results (i.e., results showing positive, statistically significant results). As acknowledged elsewhere (Young and Kindzierski 2019), search space counts are considered lower bound approximations of the number of test possible.

Creswell (2003) indicates that a 5−20% sample from a population whose characteristics are known is considered acceptable for most research purposes as it provides an ability to make generalizations for the population. Given the prior screening and data collection procedures used by Zheng et al., we assumed their 87 papers had consistent characteristics suitable for use in meta-analysis. Based on this assumption, we randomly selected 17 of the 87 papers (~20%) for search space counting in the following manner. Using similar methods described previously (Young et al. 2019; Young and Kindzierski 2018, 2019), we started with the 87 base papers and assigned a separate number in ascending order to each paper (numbered 1−87). We then used the online web tool numbergenerator.org to generate 10 random numbers between 1 – 87. We then removed the 10 selected papers from the ordered list, renumbered the remaining papers 1−77 and used the web tool to generate 7 random numbers between 1 – 77. This allowed us to select 17 of the Zheng et al. base papers for further evaluation (refer to SI 2).

Electronic copies of the 17 randomly selected bases papers (and any corresponding electronic supplementary information files) were obtained and read. One change was made from previous search space counting procedures used by Young et al. (2019) and Young and Kindzierski (2019). It was apparent that several of the base papers employed a variety of model forms in their analysis (e.g., Sheppard et al. 1999, Lin et al. 2002, Tsai et al. 2006, Hua et al. 2014). To accommodate this, we separately counted the number of model forms along with the number of outcomes, predictors, time lags, covariates reported in each base paper (covariates can be vague as they might be mentioned anywhere in a base paper). Specifically, analysis search space of a base paper was estimated as follows:

- The product of outcomes, predictors, model forms and time lags = number of questions at issue, Space1.
- A covariate may or may not act as a confounder to a predictor variable and the only way to test for this is to include/exclude the covariate from a model. As it can be in or out of a model, one way to approximate the modelling options is to raise 2 to the power of the number of covariates, Space2.
- The product of Space1 and Space2 = an approximation of analysis search space, Space3.

Table 2 provides three examples of search space analysis of a hypothetical observational study of ambient air quality versus hospitalization due to asthma exacerbation.

*p-Value plots*

p-Value plots following the ideas of Schweder and Spjøtvoll (1982) were then developed to inspect the distribution of the set of p-values reported for each ambient air quality component in the Zheng et al. meta-analysis. The p-value can be defined as the probability, if nothing is going on, of obtaining a result equal to or more extreme than what was observed. The p-value is a random variable derived from a distribution of the test statistic used to analyze data and to test a null hypothesis. Under the null hypothesis, the p-value is distributed uniformly over the interval 0 to 1 regardless of sample size (Hung et al. 1997) and a distribution of true null hypothesis points in a p-value plot should form a straight line (Schweder and Spjøtvoll 1982).

A plot of p-values sorted by rank corresponding to true null hypothesis points should conform to a near 45 degree line. The plot can be used to assess the validity of a false claim being taken as true and, specific to our interest, can be used to examine the reliability of base papers used in the Zheng et al. meta-analysis. P-value plots for each ambient air quality component were constructed and interpreted as follows (after Schweder and Spjøtvoll 1982):

- P-values were computed using the method of Altman and Bland (2011) and ordered from smallest to largest and plotted against the integers, 1, 2, 3, …
- If the points on the plot followed an approximate 45-degree line, then the p-values are assumed to be from a random (chance) process.
- If the shape of the points exhibits a hockey stick, i.e., bilinear shape (blade on the bottom left hand corner, shaft towards the top right hand corner), then the p-values used for meta-analysis constitute a mixture and a general (over-all) claim is not supported; in addition, the p-value reported for the overall claim in the meta-analysis paper is not valid.

To assist in the interpretation of behavior of the Zheng et al. meta-analysis p-value plots, we also constructed and show p-value plots for plausible true null and true alternative hypothesis outcomes based on meta-analysis of observational datasets.

**Results**

*Analysis search space*

Estimated analysis search spaces for the 17 randomly selected base papers from Zheng et al. are presented in Table 3. From Table 3, investigating multiple (i.e., 2 or more) asthma outcomes in the selected base were as common as single outcome investigations. In addition, use of multiple models and lags was common in their analysis, so was making adjustments for multiple possible covariate confounders. While the multiple factors considered (i.e., outcomes, predictors, models, lags and treatment of covariates) is seemingly realistic, these attempts to find possible exposure−disease associations among combinations of these factors will increase the overall number of tests performed in a single study.

Summary statistics of the possible numbers of tests in the 17 base papers are presented in Table 4. The median number of possible statistical tests (Space3) of the 17 randomly selected base papers was 15,360 (interquartile range 1,536 – 40,960), in comparison to actual statistical test results presented. Given these large numbers, statistical test results taken from the base papers are unlikely to provide unbiased measures of effect for meta-analysis.

*p-Values*

Altman and Bland (2011) indicated that their method for calculating p-values is unreliable for very small values and to report as p<.0001 if a calculated p-value is smaller than .0001. Consequently, p-values calculated as <.0001 were reported simply as .0001. This was done in order to facilitate creation of p-value plots for each air quality component. Zheng et al. drew upon many (332) summary statistics from 87 base papers for their meta-analysis of six air quality components. Summary statistics they used in their meta-analysis of PM2.5 (i.e., Risk Ratio (RR), Lower Confidence Level (LCL) and Upper Confidence Level (UCL) values) and along with p-values calculated after Altman and Bland (2011) are presented in Table 5. Summary statistics for the other air quality components (CO, NO2, O3, SO2 and PM10) and calculated p-values are provided in SI 3. In Table 5 (and tables presented in SI 3), calculated p-values ≤.05, taken as a statistically significant result, are bolded and italicized.

Table 6 presents additional information on each of the air quality components (i.e., RR counts and p-value summaries). Zheng et al. drew upon 37 summary statistics from 32 base papers for their PM2.5 meta-analysis. The majority of PM2.5 summary statistics had p-values greater than .05 (23 of 37), 14 of 37 were smaller than .05 and 8 of 37 were smaller than .001. Researchers often accept a p-value of .001 or smaller as virtual certainty (Young and Kindzierski 2019). If a summary statistic has a p-value small enough to indicate certainty, one should expect few p-values larger than .05 (Boos and Stefanski 2011, Johnson 2013). This is not the case as 45% or more of all summary statistics used by Zheng et al. had p-values greater than .05 (refer to Table 6).

*p-Value plots*

A plot of ranked p-values versus integers for a dataset of true null hypothesis would present as a sloped line going left to right at an approximate 45-degrees. Whereas a plot of ranked p-values versus integers for a dataset of true alternative hypothesis should have a majority of p-values in the plot should be smaller than .05.

Figure 2 presents p-value plots of meta-analysis datasets (n<13 base papers) for plausible true null hypothesis and true alternative hypothesis taken from a meta-analysis of selected cancers in petroleum refinery workers after Schnatter et al. (2018). Data used for Figure 2 are summarized in SI 4. The distribution of the p-value under the alternative hypothesis – where the p-value is a measure of evidence against the null hypothesis – is a function of both sample size and the true value or range of true values of the tested parameter (Hung et al 1997).

In Figure 2, both p-value plots are based on small sample sizes <13. However, both plots conform to that explained above – i.e., sloped line from left to right at an approximate 45-degrees for data suggesting a plausible true null chronic myeloid leukemia causal relationship in petroleum refinery workers (left image) and a majority of p-values below the .05 line for data suggesting a plausible true mesothelioma causal relationship in petroleum refinery workers (right image).

Figure 3 presents another set of p-value plots of meta-analysis datasets for large numbers of base papers (n>65) showing plausible true null hypothesis and true alternative hypothesis. Data used for Figure 3 are summarized in SI 5 and SI 6. Figure 3 plots also conform to expected behaviour, similar to Figure 2 – sloped line from left to right at an approximate 45-degrees for data suggesting a plausible true null hypothesis for elderly long-term exercise training−mortality & morbidity risk (after de Souto Barreto et al. 2019) (left image) and a majority of p-values below the .05 line for data suggesting a plausible true alternative hypothesis for smoking−squamous cell carcinoma risk (after Lee et al. 2012) (right image).

Figure 4 presents p-value plots for each of the six air components in the Zheng et al. meta-analysis. All of these plots are different from p-value plot behavior of both plausible true null and true alternative hypothesis outcomes (Figures 2 and 3). Variability among data is apparent in p-value plots for each air component. For example, the p-value plot for PM2.5 (lower left image) presents as a distinct two-component mixture resembling a hockey stick– the blade (small p-values which may suggest a true positive causal relationship or a false positive causal relationship due to p-hacking) and the shaft (p-values that are consistent with random outcomes). As p-value plots are standard technology, a two-component mixture of p-values indicates a mix of studies suggesting an association and no association. Both cannot be true. The two-component shapes of the p-value plots appear consistent with the possibility of analysis manipulation to obtain small p-values in several of the base papers. These p-value (mixture) relationships do not support a real exposure−disease (air quality−asthma exacerbation) claim.

**Discussion**

*Asthma characteristics*

Asthma is a disease that affects the airways to the lungs and it is worthwhile summarizing what is presently known about the disease as it relates to relevant characteristics, genetics and socioeconomic factors, triggering/precipitating factors and seasonality patterns of exacerbations.

<u>Development of asthma</u> – Asthma is associated with three principal characteristics (Pillai and Calhoun 2014, Noutsios and Floros 2014): (1) variable airways obstruction, (2) airway hyperresponsiveness, and spasm with wheezing and coughing and (3) airway inflammation. Clinically asthma is characterized by episodic, reversible obstructive airways obstruction that variably presents as a variety of symptoms from cough to wheezing, shortness of breath, or chest tightness (Busse and Lemanske 2001).

The underlying pathology of asthma, regardless of its severity, is chronic inflammation of the airways and reactivity/spasm of the airways. A combined contribution of genetic predisposition and non-genetic factors account for divergence of the immune system towards T helper (Th) type 2 cell responses that include production of pro-inflammatory cytokines, Immunoglobulin E (IgE) antibodies and eosinophil infiltrates (circulating granulocytes) known to associate with asthma (Figure 5) (Noutsiosa and Floros 2014). The release of pro-inflammatory cytokines that cause airway narrowing is responsible for cough, shortness of breath, wheezing and chest tightness characteristic of the asthmatic state (Bernstein 2008). But this fails to account for beta stimulus of bronchiolar muscles that increases airway spasm. Airway inflammation causes secretions and contributes to edema (swelling) but it does not cause hyperresponsiveness.

During exacerbations these airway narrowing processes are accentuated, but it is not completely clear how these events contribute to these underlying changes and the mechanisms underlying an increase in airflow obstruction are not fully understood (Singh and Busse 2006). There is also ongoing debate as to whether asthma is one disease or several different diseases that include airway inflammation; however two thirds (or more) of asthmatic patients have an allergic component to their disease and are felt to have allergic asthma (Scherzer and Grayson 2018). Not enough is currently known to rule out allergic causes for a vast majority of asthma problems. As for development of the disease, asthma frequently first expresses itself early in the first few years of life arising from a combination of host and other factors (Table 7). Most investigators would agree there is a major hereditary contribution to the underlying causes of asthma and allergic diseases (Lemanske and Busse 2003).

*Triggering/precipitating factors* – Triggering/precipitating factors that may provoke exacerbations or continuously aggravate symptoms (secondary causes of asthma) are summarized in Table 7. Respiratory viruses are present in a majority of asthmatic children during episodes of exacerbation (Johnston 2007, Szefler 2008, Costa et al. 2014). Airways are made more reactive by infection, so asthmatic children have episodes of asthma that are aggravated or initiated by respiratory infections.

Mechanisms by which air quality possibly contributes to asthma exacerbation in subjects with existing airway allergies are thru immune responses, including:

- Producing Immunoglobulin E antibody responses in the immune system – alteration of T regulatory cell function (which has a role in regulating or suppressing other cells in the immune system) and changes in $FEV_1$ (forced expiratory volume in 1 second) (Diaz-Sanchez 1999, Nadeau et al. 2010). However, changes in $FEV_1$ are only a measure of airway obstruction and not evidence of a causal factor.
- Modulating or activating inflammasomes (Bauer et al. 2012). Inflammasomes are cytosolic multiprotein oligomers of the innate immune system responsible for activation of inflammatory responses (Latz et al. 2013). They are key signalling platforms that detect pathogenic microorganisms and sterile stressors, and that activate highly pro-inflammatory cytokines interleukin-1β (IL-1β) and IL-18.
- Impairing defences of the airway epithelium through reduced barrier function, impaired host defence to pathogens, and inflammatory responses from generation of reactive oxygen species (ROS) and resultant oxidative stress (Huang et al. 2015, De Grove et al. 2018, Huff et al. 2019).

From an exposure point-of-view, isolating the role/contribution of a particular triggering/precipitating factor in an asthma exacerbation episode is difficult, unless the factor overwhelms. As asthma is a complex interaction between the inhaled environment and the formed elements of the airways (Holgate 2010), the hypothesis is that exposure to abruptly changing air quality conditions may contribute to symptoms and increase the severity of asthma exacerbations; although these effects are not as pronounced as those of viruses and aeroallergens (Trasande and Thurston 2005). By the examination of multiple air quality component−asthma exacerbation observational studies, our study helps judge if this hypothesis is supported.

Because of multiple triggering/precipitating factors that may be at play, D'Amato et al. (2005) indicate it is challenging to evaluate the role of air quality on the timing of asthma exacerbations and on the prevalence of asthma in general. For example, enhanced Immunoglobulin E antibody-mediated response to aeroallergens and resulting enhanced airway inflammation could account for increasing frequency of allergic respiratory allergy and bronchial asthma often attributed to air quality components in observational studies.

*Seasonality*– There is seasonal variation in hospitalization of children due to asthma exacerbations, with fewer hospitalizations in summer and more in fall and winter in North America (e.g., United States – Kimes et al. 2004, Sokol et al. 2014; Canada – Crighton et al. 2001, Johnston 2007). As for adults (≥50 years), Johnston (2007) reports that a major asthma exacerbation epidemic occurs annually during the Christmas period, and hovers near average levels for the remainder of the year other than a mild peak in early spring.

International studies reveal the same seasonal variation in asthma exacerbations and peak hospitalizations of children in fall – England (Khot et al. 1984, Jackson et al. 2011), Netherlands (Koster et al. 2011), Finland (Harju et al. 1997), Norway (Carlsen et al. 1984), Malta (Grech et al. 2002), New Zealand (Kimbell-Dunn et al. 2000), Israel (Garty et al. 1998), Taiwan (Xirasagar et al. 2006) and Hong Kong (Tseng et al. 1989).

Asthma exacerbations are not so problematic in summer except for those asthmatics that are triggered by seasonal allergens that appear in summer. The onset of cooler weather brings on asthma exacerbations. Wintertime allergenic stimulants is more a problem of inside air—insect offal and dander and household allergens as well as mold in closed in housing or other allergens that are accentuated by inside living. There is also the well-known asthmatic trigger of cooler or cold air.

Possible causes of higher rates of asthma hospitalizations in fall proposed by others include climate factors, increased incidence of viral infections and changes in air quality components (Xirasagar et al. 2006, Jackson et al. 2011, Koster et al. 2011, Sokol et al. 2014). However, house mold and dander as well as insect offal and detritus may also be at play. Crighton et al. (2001) stated that ambient levels of air pollutants vary significantly from place to place, and in Ontario Canada these levels typically peak during summer months when asthma hospitalizations are at their lowest. Warmer summer temperatures typically result in an increase in levels of ozone precursors and ozone yet asthma hospitalizations decline. Rather than there being a number of single spiked events such as abrupt changes (increases) in air quality components on a given day, there is more likely to be other triggering/precipitating factors that accumulate and diminish over many days to weeks that more plausibly explain asthma hospitalizations of children in the fall in North America (Kimes et al. 2004).

*Multiple testing multiple modelling (MTMM) problem*

The median (interquartile range) analytical search space count of 17 randomly selected base papers reviewed is considered large – 15,360 (1,536 – 40,960) – indicating the potential of large numbers of statistical tests conducted in the base papers. We also noted that search space counts of air quality component−heart attack observational studies in published literature are similarly large (i.e., median (interquartile range) = 6,784 (2,600 − 94,208), n=14 (Young et al. 2019) and 12,288 (2,496 − 58,368), n=34 (Young and Kindzierski (2019).

The MTMM problem can be explained in this way – MTMM involves researchers performing many statistical tests; whereas analysis manipulation involves researchers only selecting and reporting a portion of their findings – ones with positive associations – in a study, possibly to support a point-of-view. Researchers then only need to describe research designs and methods in their studies consistent with their reported findings (and point-of-view) and other findings, e.g., negative associations, tend to be ignored.

This can be written up in a professional manner and submitted to a scientific journal. Journal editors can overlook MTMM practices given the professional, tight presentation of a scientific manuscript and send it off for independent review; likewise for independent reviewers. In the end what gets published can be based on a portion of the tests actually performed (i.e., selectively reported test results) and disregarding whether these are true or false positive findings. Others (Feinstein 1988, Freedman 2010, Keown 2012) have reported that studies like these are more likely presenting false positive findings.

For any given set of multiple null hypothesis tests, 1 of every 20 p-values could be .05 or less even when the null hypothesis is true based on the Neyman-Pearson theory of hypothesis testing – known as the Type I (false-positive) error rate (Hung et al. 1997, Lew 2020). Put another way, a p-value less than .05 does not by any means indicate that a positive outcome is not a chance finding (Bock 2016). When many null outcomes are tested, the expected number of false positive outcomes can be inflated compared to what one might expect or allow due to chance (typically 5%) as the overall chance of a false positive error can exceed the nominal error rate used in each individual test (Parker and Rothenberg 1988, Westfall and Young 1993, Shaffer 1995, Young and Kindzierski 2019, Makin and de Xivry 2019). Large numbers of null hypothesis tests were a common feature of the 17 randomly selected base papers reviewed. Thus for these base papers, the upper limit of expected number of false positive outcomes may exceed 5% of the tests performed.

The problem may worsen with MTMM of observational datasets whose explanatory (predictor) variables are not independent of each other. For example, in urban areas where most observational air quality–effect studies draw data from, there are known correlations between air quality components related to co-pollutant emissions from motor vehicle, commercial and industrial activities – e.g., $NO_2$ with CO and $SO_2$; $PM2.5$ with $PM10$, $NO_2$, CO and $SO_2$, and $SO_2$ with CO (Sheppard et al. 1999, Lin et al. 2002, Cheng et al. 2014). Here, the expected overall false positive error rate can decrease or become inflated. Any variable correlated with a false positive variable may also be selected. But the effective number of variables will be reduced.

On balance correlations will reduce the number of claimed associations, but they will add to the complexity even more among results when multiple tests (comparisons) of co-dependent explanatory variables are made (Royall 1986, Bland and Altman 1995, Bender and Lange 2001, Ilakovac 2009, Graf et al. 2014). In addition, Simmons et al. (2011) showed that employing a few common forms of p-hacking can cause the false positive error rate for a single study to increase from the nominal 5% to over 60%.

One is unable to conclude anything of consequence by observing 1 positive (statistically significant) result from 20 independent statistical null hypothesis tests based on a 5% false positive error rate. If these tests are not independent and p-hacking is employed, one is unable to conclude anything of consequence even observing more than 1 positive result because of the potential for false positive error rate inflation. The probability of this occurrence depends on a host of factors and is almost never uniform across the tests performed (thus violating a key assumption of the 1 in 20 error rate rule of a null hypothesis test). If statistical null hypothesis testing is used as a kind of data beach-combing tool unguided by clear (and ideally prospective) specification of what findings are expected and why, much that is nonsense will be "discovered" and added to the peer-reviewed literature (Mark et al. 2016). Makin and de Xivry (2019) states that failing to correct for MTMM can be detected by estimating the number of independent variables measured and the number of analyses performed in a study – which is why we estimated analysis search spaces (i.e., number of analyses performed).

The MTMM problem, as stated previously, is analogous to 'Model Selection' in statistical models used to estimate the effects of environmental impacts (Koop and Tole 2004, Koop et al. 2010). Given $k$ potential explanatory variables for a model, a researcher can in principle run $2^k$ possible regressions using Model Selection and select one of the models that yields the 'best' results. Coefficient standard errors for this model will be invalid because it does not take into account model uncertainty. Clyde (2000) and Koop et al. (2007, 2010) instead recommended Bayesian Model Averaging (BMA), which entails estimating all $2^k$ models then generating posterior distributions of the coefficients weighted by the support each model gets from the data.

Of interest to our study, Koop et al. (2007) used both Model Selection and BMA techniques for estimating statistical associations between air quality component levels and two respiratory health outcomes – i) monthly hospital admission rates by age group and ii) length of patient days in hospital – for all lung diagnostic categories, including asthma, in 11 large Canadian cities from 1974 to 1994. Koop et al. (2007) observed that when using hospital admission rates as the dependent variable, only insignificant or negative effects were estimated regardless of whether Model Selection and BMA were used. When using patient days as the dependent variable, Koop et al. (2007) observed that none of the air quality components showed a significant positive effect on health, again regardless of whether Model Selection and BMA was used.

These findings are particularly notable given that the time period of their analysis – 1974 to 1994 – represented conditions with much higher air quality component levels in Canadian cities then compared to today. Koop et al. (2007) also highlighted the danger that incomplete modelling efforts – such as using Model Selection techniques – could yield apparent air quality component−health associations that are not robust to reasonable variations in estimation methods.

*Lack of transparent descriptions of statistical tests and statistical models*

In reviewing the 17 randomly-selected based papers, transparency was lacking in the methodology descriptions making it challenging for us and, in general, readers to understand how many statistical tests and models were used in these studies. Consequently our reviews required careful reading and re-reading of the 'Method' section, and the 'Results' and "Discussion' sections of the base papers to understand what was actually done and to compile information for estimating the analysis search spaces. As an example, in one base paper reviewed (Evans et al. 2013) only near the end of their Discussion section did they indicate that multiple pollutants were included in the same model – implying that multivariate models were employed along with univariate and bivariate models that they described in their Methods section.

Why this lack of transparency is important is explained further. It is common today for observational studies employing MTMM to seek out information on multiple exposures and disease outcomes, and the possibility exists for researchers to test thousands of exposure–disease combinations (Boffetta et al. 2008) and only report a portion of results that allow them to make interesting claims. Several hypothetical analytical search space scenarios were presented in Table 2. A single model form was used for these scenarios, although in practice it is not uncommon to use multiple modelling forms (e.g., Clyde 2000, Sheppard et al. 1999, Lin et al. 2002, Tsai et al. 2006, Hua et al. 2014).

As shown in Table 2, analysis search spaces can easily inflate into tests of thousands of exposure–disease combinations. Without clear, concise descriptions about details of outcomes, predictors, models, lags and covariate confounders used (or available) in a study, readers will not be able to comprehend what was done relative to the few statistical results that typically get presented in a study. The latter Examples 2 and 3 in Table 2 illustrate opportunities for researchers to search through but only report statistically significant, positive exposure–disease associations in a study.

Selective reporting of only these findings is not meaningful in the face of false positive error rates for a large number of tests. Specifically, theoretical expected numbers of false positives within a large number of tests, based on a false positive error rate of 5% is = Space3 x 5% (independent tests) and $\geq$ Space3 x 5% (for correlated tests) (Bland and Altman 1995, Bender and Lange 2001, Ilakovac 2009). Given the dependency of numerous variables test in the models, discussed above, expected numbers of false positive findings for Examples 2 and 3 from Table 2 could be:

- Example 2 – $\geq$2,304 x 0.05 (i.e., $\geq$115)
- Example 3 – $\geq$23,040 x 0.05 (i.e., $\geq$1,152)

Performing large numbers of statistical tests without offering all findings (which is now possible with supplemental material and web posting), and how and whether the dependency issue is dealt with among known correlated explanatory variables tested makes it challenging to ascertain how many true or false positive versus negative findings might exist in an analysis. The issue of whether p-hacking was used also cannot be disentangled when only a few of the findings are presented. The reader cannot make a reasoned judgment decision about the few statistically significant findings that are presented.

*Heterogeneity*

Any kind of variability among studies in a systematic review or meta-analysis may be termed 'heterogeneity' (Higgens and Green 2008). It arises because of effects in the populations which the studies represent are not the same or even the analysis methods employed are not the same. Beyond the role of chance, the influence of bias, confounding or both can contribute to heterogeneity among observational studies (Egger et al. 1998).

The median number of possible statistical tests of the 17 randomly selected base papers from Zheng et al. was large – a median of 15,360 (interquartile range 1,536 – 40,960). Further, p-value plots of the six air quality components (Figure 3) show a two-component mixture – i.e., obvious variation – common to all components with small p-values used from some base papers (which may largely be due to p-hacking) while other p-values appear random (i.e., >.05). In this case, estimating an overall effect (association) in meta-analysis by averaging over the mixture does not make sense. The p-value plots in Figure 4 do not directly confirm evidence of p-hacking in the base papers; however, in the presence of a large number of statistical tests in the base papers reviewed (Tables 2 and 3), p-hacking cannot be ruled out as an explanation for small p-values.

Heterogeneity will always exist whether or not one can detect it using a statistical test (Higgens and Green 2008). Zhang et al. calculated statistical heterogeneity ($I^2$) – which quantifies the proportion of the variation in point estimates due to among−study differences. For their overall analysis of the six air components, they reported $I^2$ values of 85.7% (CO), 69.1% (PM10), 82.8% (PM2.5), 77.1% (SO2), 87.6% (NO2) and 87.8% (O3).

Higgens et al. (2003) assigned low, moderate and high $I^2$ values of 25%, 50%, and 75% for meta-analysis. Higgens and Green (2008) provide another guide to interpretation: 0−40% might not be important, 30−60% may be moderate heterogeneity, 50−90% may represent substantial heterogeneity and 75−100% represents considerable heterogeneity. These criteria suggest that meta-analysis of all air components (except PM10) by Zheng et al. are associated with high/substantial heterogeneity (i.e., >75%).

Forstmeier et al. (2017) note that a key source of heterogeneity in meta-analysis is publication bias in favor of positive effects, often facilitated by use of researcher degrees of freedom (flexibility) to find a statistically significance effect more often than expected by chance. Zheng et al. acknowledge that publication bias was detected in all of their analyses except for PM2.5. In addition, Zheng et al. – as described in their Methods section – indicated the screening and data collection procedures for identifying their base studies complied with PRISMA – Preferred Reporting Items for Systematic Reviews and Meta-analysis. PRISMA is a 27-item checklist to use for reporting items for systematic reviews and meta-analyses (Moher et al. 2009). PRISMA is one of dozens of quality appraisal methods that exist in research synthesis studies (Wells and Littell 2009); however we note that the PRISMA checklist is silent on MTMM (and possible p-hacking). Publication bias and MTMM may, in part, explain the nature of the heterogeneity – i.e., two-component mixture (refer to Figure 4). These factors cannot be dismissed as possible explanations for their findings.

*Limits of observational epidemiology*

All of the Zheng et al. 17 randomly selected base papers that we reviewed made use of exposure−response models that, in reality, cannot address the complexities related to real-world air quality component exposure−asthma responses. These models represented the population level and did not capture individual level behaviors and possible exposures to other triggers and other real-world confounders. Admittedly, these are difficult to capture in population-based studies due to feasibility and cost limitations.

An important criterion supporting causality of an air quality component−asthma exacerbation association is a dose−response relationship. In the absence of flexibility in data collection, analysis and reporting available to researchers, exposure to a true trigger cannot both '*cause*' and '*not cause*' asthma exacerbation per equivalent unit change in air quality component level (i.e., per 10 per $\mu g/m^3$ increase) across populations tested. Of the 332 Zheng et al. summary statistics (RRs) used for meta-analysis of air quality components (Table 5), 196 (59%) represented null (statistically non-significant) associations. That is to say, they offer insufficient evidence to support an air quality component exposure−asthma exacerbation causal relationship.

Quantitative results from observational studies (e.g., RRs, odds ratios) can figure prominently into regulatory decisions but frequently observational studies offer RRs and odds ratios extremely close to 1.0. Further, a disservice to observational epidemiology is the practice of searching for and reporting and attempting to defend weak statistical associations (e.g., RRs and odds ratios extremely close to 1.0) – among which the potential for distorting influences of chance, bias and confounding is further enhanced (Boffetta et al. 2008). In its simplest form, risk factor−chronic disease observational epidemiology examines statistical relationships (associations) between risk factors (independent variables) and chronic diseases (dependent variables) in the presence/absence of confounders and moderating factors (covariates) that may or may not alter these relationships. However, observational associations between variables do not guarantee causality and they are often complex and influenced by other variables – e.g., confounders and moderating factors (Palpacuer et al. 2019).

Biomedical literature is largely absent on evidence that air quality components represent a strong risk factor to the common chronic diseases in United States unlike evidence for factors such as unhealthy diet, physical inactivity, tobacco use and harmful substance abuse (e.g., alcohol) (Beaglehole et al. 2007, Alwan and MacLean 2009). For air quality component−chronic disease observational studies to offer meaningful statistical results and meaningful evidence for public health practitioners, risk results (e.g., RRs, odds ratios) need to be high enough, whether in the presence/absence of confounders and moderating factors, to be taken seriously. For example, Ahlbom et al. (1990), Beaglehole et al. (1993) and Federal Judicial Center (2011) recommend RRs>2 to rule out bias and confounding. Of 332 RRs used in the Zheng et al. meta-analysis for all six air quality components, only 2 (<1%) had RRs >2. Thus we are unable to rule out bias and confounding out as explanations for the vast majority (>99%) of RRs used for meta-analysis by Zheng et al. from the 87 base papers.

Epidemiology studies that test many null hypotheses tend to provide results of limited quality for each association due to limited exposure assessment and inadequate information on potential confounders (Savitz and Olshan 1995) and they tend to generate more errors of false positive or false negative associations (Rothman 1990). We view the Zheng et al. meta-analysis as a possible example of what is known as 'vibration of effects' after Ioannidis (2008b), Patel et al. (2015) and Palpacuer et al. (2019) – wherein summary statistics for a weak risk factor (i.e., ambient air quality) are combined from base studies whose outcomes vibrate between suggesting associations/no associations dependent upon how researchers designed the studies, collected and analysed data and reported their results. Flexibility in data collection, analysis and reporting available to researchers dramatically increases actual false-positive rates (Simmons et al. 2011) and this was not considered by Zheng et al. as a possible explanation for their findings. Our findings do not support an ambient air quality component−asthma exacerbation causal relationship.

*Recommendations for improvement*

It is our belief that risk factor−chronic disease researchers are unaware that many positive research findings from published observational studies may be false. Also, there are many sources of bias currently being underestimated by observational study researchers with selective reporting biases likely a key issue distorting their findings, and publication bias is likely a key issue distorting the epidemiologic literature in general. As a result, we present a number of recommendations for ways of improving risk factor−chronic disease observational studies to address these issues.

Our recommendations have largely been advanced by others in the past– e.g., Simes (1986), Begg and Berlin (1988), Angell (1989) and Dickersin (1990). In addition, this issue has become a topic of interest more recently – e.g., Song et al. (2010), Lash and Vandenbroucke (2012), Wicherts et al. (2012), Connell et al. (2014), Vandenbroucke et al. (2014), Wicherts et al. (2016), Lakens et al. (2016), Wicherts (2017), Forstmeier et al. (2017), Randall and Welser (2018), NASEM (2019) and Miyakawa (2020).

In regards to individual researchers improving risk factor−chronic disease observational studies, it is apparent to us that all roads pass thru funding agencies and journal editors (and reviewers). Many funding agencies and journals largely emphasize novelty of research for publication (Wicherts 2017, Forstmeier et al. 2017). Researchers are aware of this bias to the novel. Regrettably, many researchers are also aware of the possibility of *you can publish if you find a significant (positive) effect* (Forstmeier et al. 2017). This motivates researchers to engage in MTMM and other questionable research practices (Banks et al. 2016) to produce and publish positive research findings in order to advance their careers.

NASEM (2019) recently reported that a survey of researchers found that researchers themselves believe they are responsible for addressing issues of reproducibility, but that a supportive institutional infrastructure (e.g., training, mentoring, funding and publishing) is needed. While the importance of this cannot be dismissed, we expect there will be little incentive to change current practices at the individual researcher-level unless there is a commitment to change how funding agencies, journal editors (and reviewers) conduct their end of the complex science industry publication business.

Therefore our recommendations are aimed specifically at funding agencies and journal editors. Although, many useful recommendations exist in literature for individual researchers (e.g., refer to Song et al. 2010, Gelman and Loken 2013, Lakens et al. 2016, Wicherts et al. 2016, Forstmeier et al. 2017, Wicherts 2017, NASEM 2019, Makin and de Xivry 2019, Miyakawa 2020).

We see the following topics as being crucial for enabling improvements in risk factor−chronic disease observational studies at the funding agency/journal level:

- Preregistration.
- Changes in funding agency, journal editor (and reviewer) practices.

- Open sharing of data.
- Facilitation of reproducibility research.

*Preregistration* – Preregistration involves defining the research hypothesis to be tested, identifying whether the study is confirmatory or exploratory and defining the entire data collection and data analysis protocol prior to conducting an observational study (Lash and Vandenbroucke 2012, Lakens et al. 2016, Wicherts 2017, Forstmeier et al. 2017, NASEM 2019). Preregistration is intended to address issues of (Forstmeier et al. 2017): (i) hypothesizing after results are known (HARKing), (ii) minimize researcher flexibility in data analysis and reporting of findings (p-hacking) and (iii) publication bias.

Wicherts (2017) proposed that preregistration should be openly published, time-stamped (and should not be changed), and be sufficiently detailed to establish that a study was actually done in line with the hypothesis and free from biases due to how researchers analyze the data and report the results. Connell et al. (2014) notes that it is crucial for descriptions of data collection and analysis protocols to be transparent; for example, to the point where readers have a clear understanding of details about the number of independent variables measured and the number of analyses performed in an observational study (Makin and de Xivry 2019). This was a key limitation that we faced in our present evaluation of the Zheng et al. meta-analysis.

*Changes in funding agency, journal editor (and reviewer) practices* – Funding agencies and research sponsors tend to emphasis novelty of research (Forstmeier et al. 2017). These efforts should be reallocated towards supporting replication of important research findings for the benefit of the scientific community as a whole. Concurrent with this is funding agencies requiring preregistration and open sharing of data (i.e., making data available to the pubic) as a requirement for funding observational studies.

Journals too seek novelty in research in part because of the competition for impact factors (Forstmeier et al. 2017). In fact, editors are often rewarded for actions that increase the impact factor of the journal (NASEM 2019). Studies which report novel findings are more often highly cited and thus contribute to the stature of a journal. This should not be a focus. Rather, journals and journal editors should formalize editorial policies and practices around making decisions to publish based on issues of quality and logical reasoning by researchers and not on novelty or the direction (i.e., positive versus negative results) and strength of study results (Dickersin 1990).

*Open sharing of data* – We consider the potential for bias to be particularly severe for observational studies that investigate subtle/weak risk factor−chronic disease relationships. This type of epidemiological research poses a considerable challenge for the most common chronic diseases of interest in United States (Table 1) because this research operates in areas with very low pre- and post-study probability for true findings (Figure 1). Along with preregistration, another way to deal with this challenge is to make the data from these studies openly available for independent reanalysis after publication (Wicherts et al. 2017, Randall and Welser 2018, NASEM 2019, Miyakawa 2020).

*Facilitation of reproducibility research* – Funding agencies and journals are seldom willing to fund or allow publication of replication research, particularly when the results contradict earlier findings (Wicherts 2017). As replication research fulfils an important need in the advancement of scientific knowledge and to address publication bias (Lakens et al. 2016, Wicherts 2017), funding agencies and journals should fund and publish replication research.

**Summary**

False-positive results and bias may be common features of the biomedical literature today, including risk factor−chronic disease research. Included with this are observational studies of ambient air quality as a chronic disease risk factor, which are known to operate in areas with low probability of positive findings being true for the most common chronic diseases, including asthma, in United States. Because of these potential problems, we undertook an evaluation of the reliability of observational base studies used in the Zheng et al. meta-analysis examining whether six ambient air quality components trigger asthma exacerbation. We observed that the median number of possible statistical tests of 17 randomly selected base papers from Zheng et al. was large – 15,360 (interquartile range 1,536 – 40,960) – suggesting that large numbers of statistical tests were a common feature of the bases studies.

Given this, p-hacking cannot be ruled out as explanations for small p-values reported in some of the base papers. We also observed that p-value plots of the six air quality components showed a two-component mixture common to all components – with p-values from some base papers having small values (which may be false-positives) while other p-values appearing random (>.05). The two-component shapes of these plots appear consistent with the possibility of analysis manipulation to obtain small p-values in several of the base papers.

Regarding two properties of the Zheng et al. meta-analysis that we were interested in understanding:

- As for the reliability of claims in the base papers of their meta-analysis due to the presence of multiple testing and multiple modelling – which can give rise to false positive results, we conclude that the meta-analysis is unreliable due to the presence of multiple testing and multiple modelling.
- As for whether heterogeneity across the base papers of their meta-analysis is more complex than simple sampling from a single normal process, we show that the two-component mixture of data used in the meta-analysis (i.e., Figure 4) does not represent simple sampling from a single normal process.

Our interpretation of the Zheng et al. meta-analysis is that the random p-values indicating null associations are more plausible and that their meta-analysis will not likely replicate in the absence of bias. We conclude that the Zheng et al. meta-analysis and the corresponding base papers they drew upon are unreliable and do not offer evidence of value to inform public health practitioners about ambient air quality as a risk factor for asthma exacerbation.

It is our belief that risk factor−chronic disease researchers are unaware many positive research findings from published observational studies may be false. In this regard, we see the following areas as being crucial for enabling improvements in risk factor−chronic disease observational studies at the funding agency and journal level: preregistration, changes in funding agency, journal editor (and reviewer) practices, open sharing of data and facilitation of reproducibility research.

**Acknowledgements**

This study was developed based on work originally undertaken for the National Association of Scholars (nas.org), New York, NY. The authors gratefully acknowledge Gerhard Benade, Jerry Arnett and Lawrence Davis for helpful discussions and feedback on the writing about the characteristics of asthma. The authors also gratefully acknowledge comments provided by five reviewers selected by the Editor and anonymous to the authors. The comments helped improve the quality of the final article.

**Declaration of interest**

Dr. Kindzierski was an Adjunct Professor at the University of Alberta, Edmonton, Alberta, Dr. Young was self-employed with CGStat, Raleigh, North Carolina, Dr. Meyer was self-employed with Outcome Based Medicine, Raleigh, North Carolina and Dr. Dunn was retired during the writing of this manuscript. During this time all the authors have not participated in any capacity in litigation or regulatory proceedings or an advocacy role on behalf of any organization related to the contents of the paper. The research undertaken and preparation of the paper was the professional work product of all the authors. All authors have sole responsibility for the writing and content of the paper. The work has not been influenced by anyone other than the authors and the paper was not shared with any other individuals during its preparation. The analysis performed and the conclusions drawn are exclusively those of the authors. The authors report that no financial interest or benefit will arise from the direct application of this research.

**ORCID**

Warren B. Kindzierski http://orcid.org/0000-0002-3711-009X

S. Stanley Young https://orcid.org/0000-0001-9449-5478

**Supplementary material**

Supplemental information for this article is included at the end.

Koop G, Tole L. 2004. Measuring the health effects of air pollution: to what extent can we really say that people are dying from bad air? J Environ Econ Manage. 47:30−54. doi:10.1016/S0095-0696(03)00075-5.

Koop G, McKitrick R, Tole L. 2007. Does air pollution cause respiratory illness? A new look at Canadian cities. Available at https://strathprints.strath.ac.uk/7736/6/strathprints007736.pdf.

Koop G, McKitrick R, Tole L. 2010. Air pollution, economic activity and respiratory illness: Evidence from Canadian cities, 1974-1994. Environ Model Softw. 25(7):873–885. doi:10.1016/j.envsoft.2010.01.010.

Koster ES, Raaijmakers JA, Vijverberg SJ, van der Ent CK, Maitland-van der Zee AH. 2011. Asthma symptoms in pediatric patients: differences throughout the seasons. J Asthma. 48(7):694–700. doi:10.3109/02770903.2011.601780.

Labos CL, Thanassoulis G. 2018. Selection bias in cardiology research: Another thing to worry about (and how to correct for it). Can J Cardiol. 34(6):705–728. doi:10.1016/j.cjca.2018.03.010.

Lakens D, Hilgard J, Staaks J. 2016. On the reproducibility of meta-analyses: six practical recommendations. BMC Psychol. 4:24. doi:10.1186/s40359-016-0126-3.

LaKind JS, Goodman M, Makris SL, Mattison DR. 2015. Improving concordance in environmental epidemiology: A three-part proposal. J Toxicol Environ Health, Part B. 18:105–120. doi:10.1080/10937404.2015.1051612.

Lash TL, Vandenbroucke JP. 2012. Should preregistration of epidemiologic study protocols become compulsory? Epidemiol. 23(2): 184−188.

Last JM. 2001. A Dictionary of Epidemiology, 4th ed. New York, NY: Oxford University Press.

Latz E, Xiao TS, Stutz A. 2013. Activation and regulation of the inflammasomes. Nature Rev Immunol. 13:397–411. doi:10.1038/nri3452.

Lee PN, Forey BA, Coombs KJ. 2012. Systematic review with meta-analysis of the epidemiological evidence in the 1900s relating smoking to lung cancer. BMC Cancer. 12:385 (90pp.). doi:10.1186/1471-2407-12-385.

Lew MJ. 2020. A reckless guide to p-values. In: Good Research Practice in Non-Clinical Pharmacology and Biomedicine (Ed: Bespalov A, Michel MC, Steckler T). Handbook of Experimental Pharmacology Vol. 257. New York, NY: Springer. pp 223−256.

Table 1. Estimates of prevalence for common chronic diseases of interest in United States.[1]

| Disease Category | Disease | Population of interest | Prevalence rate ($P$) | Timeframe |
|---|---|---|---|---|
| Respiratory diseases | asthma | total males and females | 0.079 | in 2017 |
| | chronic obstructive pulmonary disease (COPD) | adults ≥18 years old | 0.059 | in 2014–2015 |
| Heart diseases | coronary heart disease, angina or heart attack | adults ≥18 years old | 0.056 | in 2018 |
| Diabetes | diabetes | total males and females | 0.094 | in 2015 |
| Cancers | breast | females ≥35 years old | 0.037 | in 2016 |
| | prostate | males ≥55 years old | 0.072 | in 2016 |
| | colorectal | total males and females | 0.0040 | in 2016 |
| | lung and bronchus | total males and females | 0.0017 | in 2016 |

[1]Refer to SI 1 for references upon which the prevalence rate (P) is based.

Table 2. Three examples of search space analysis of a hypothetical observational study of ambient air quality versus hospitalization due to asthma exacerbation.

*Example 1*

A simple univariate analysis of childhood asthma hospital admissions is considered using 6 air quality predictors – daily average levels of PM10, PM2.2, SO2, NO2, CO and O3, and no lags or weather covariate confounders:

- Space 1 = 1 outcome x 6 predictors x 1 model x 1 lag time (i.e., same day concentration as the event day) = 6
- Space 2 = 1 (i.e., no consideration of covariate confounding)
- Space 3 = approximation of analysis search space = 6 x 1 = 6

*Example 2*

For a simple, however slightly more typical analysis of the same 6 predictors with 3 lags (i.e., same day and 1 and 2 day lags), and 2 weather variables treated as covariate confounders (daily average temperature and relative humidity), and also adjusting for possible confounding of co-pollutants in the analysis (i.e., air quality variables are also treated as covariate confounders in the analysis), we have the following search space counts:

- Space 1 = 1 outcome x 6 predictors x 1 model x 3 lags = 18
- Space 2 = $2^{2+5}$ = 128*
- Space 3 = approximation of analysis search space = 18 x 128 = 2,304

*Example 3*

For a more typical (and more in-depth) example, four different subgroups are used in the analysis (e.g., children ≤4 years, children 5−14 years, boys only ≤14 years and girls only ≤14 years) along with the main study population:

- Space 1 = 5 outcomes (main + 4 subgroups) x 6 predictors x 1 model x 3 lags = 180
- Space 2 = $2^{2+5}$ = 128*
- Space 3 = approximation of analysis search space = 180 x 128 = 23,040

\* Note: there are 7 covariates in Space 2 – daily average temperature and relative humidity are 2 covariates; 1 pollutant is treated as a predictor adjusted with the other 5 pollutants (covariates) in the model.

Table 3. Authors, variable counts, and analysis search spaces for the 17 randomly selected base papers from Zheng et al. (2015).

| First Author | Outcomes | Predictors | Models | Lags | Covariates | Space1 | Space2 | Space3 |
|---|---|---|---|---|---|---|---|---|
| Thompson | 1 | 10 | 3 | 4 | 7 | 120 | 128 | 15,360 |
| Andersen | 3 | 11 | 1 | 6 | 8 | 198 | 256 | 50,688 |
| Chardon | 3 | 3 | 1 | 16 | 8 | 144 | 256 | 36,864 |
| Sheppard | 1 | 14 | 5 | 5 | 8 | 350 | 256 | 89,600 |
| Gouveia | 4 | 11 | 1 | 4 | 8 | 176 | 256 | 45,056 |
| Tenias | 1 | 24 | 4 | 4 | 5 | 384 | 32 | 12,288 |
| Magas | 4 | 6 | 1 | 2 | 5 | 48 | 32 | 1,536 |
| Chakraborty | 1 | 3 | 2 | 1 | 4 | 6 | 16 | 96 |
| Tsai | 1 | 10 | 2 | 3 | 2 | 60 | 4 | 240 |
| Laurent | 4 | 4 | 3 | 6 | 5 | 288 | 32 | 9,216 |
| Lavigne | 5 | 5 | 1 | 1 | 3 | 25 | 8 | 200 |
| Mar | 1 | 2 | 1 | 6 | 8 | 12 | 8 | 96 |
| Evans | 3 | 7 | 2 | 7 | 6 | 294 | 64 | 18,816 |
| Abe | 2 | 10 | 2 | 2 | 9 | 80 | 512 | 40,960 |
| Santus | 32 | 10 | 2 | 8 | 3 | 5,120 | 8 | 40,960 |
| Hua | 2 | 2 | 8 | 5 | 4 | 160 | 16 | 2,560 |
| Lin | 3 | 3 | 3 | 7 | 7 | 189 | 128 | 24,192 |

Note: Author name is first author listed (refer to SI 2).

Space 1 = Number of questions at issue = Outcomes x Predictors x Models x Lags.

Space 2 = $2^k$ where $k$ = number of Covariates.

Space 3 = Approximation of analysis search space = Space 1 x Space 2.

Table 4. Summary statistics for the number of possible tests using the three search spaces.

| Statistic | Space1 | Space2 | Space3 |
| --- | --- | --- | --- |
| minimum | 6 | 4 | 96 |
| lower quartile | 60 | 16 | 1,536 |
| median | 160 | 32 | 15,360 |
| upper quartile | 288 | 256 | 40,960 |
| maximum | 5,120 | 512 | 89,600 |
| mean | 450 | 118 | 22,866 |

Space 1 = Number of questions at issue = Outcomes x Predictors x Models x Lags.

Space 2 = $2^k$ where $k$ = number of Covariates.

Space 3 = Approximation of analysis search space = Space 1 x Space 2.

Table 5. Risk Ratio (RR), Lower Confidence Level (LCL) and Upper Confidence Level (UCL) values and corresponding p-values calculated after Altman and Bland (2011) for base papers used by Zheng et al. (2015) in their PM2.5 meta-analysis.

| Study 1st author[1] | Publication year | RR | LCL | UCL | *p-value*[2] |
|---|---|---|---|---|---|
| Lee SL | 2006 | 1.024 | 1.014 | 1.035 | ***.0001*** |
| Ko FWS | 2007 | 1.004 | 1.000 | 1.009 | .0803567 |
| Jalaludin BB | 2008 | 1.017 | 1.008 | 1.027 | ***.000432*** |
| Lavigne E | 2012 | 1.000 | 0.909 | 1.121 | 1 |
| Stieb DM | 2009 | 1.011 | 0.987 | 1.037 | .3923886 |
| Chimonas MAR | 2007 | 0.992 | 0.964 | 1.024 | .6144624 |
| Sluaghter JC (ER) | 2005 | 1.030 | 0.980 | 1.090 | .279572 |
| (H) | | 1.010 | 0.910 | 1.110 | .8548709 |
| Li S | 2011 | 1.032 | 1.007 | 1.057 | ***.010805*** |
| Mar TF | 2010 | 1.000 | 0.957 | 1.043 | 1 |
| Sheppard L | 1999 | 1.034 | 1.017 | 1.059 | ***.001249*** |
| Yamazaki S (W) | 2013 | 0.958 | 0.776 | 1.182 | .7025466 |
| (C) | | 1.039 | 0.883 | 1.222 | .6573244 |
| Santus P | 2012 | 0.991 | 0.970 | 1.011 | .399061 |
| Babin S | 2008 | 1.000 | 0.990 | 1.020 | 1 |
| Kim SY | 2012 | 1.009 | 0.991 | 1.026 | .3161553 |
| Paulu C | 2008 | 1.010 | 0.960 | 1.060 | .7070236 |
| Halonen JI (A) | 2008 | 1.003 | 0.957 | 1.050 | .907147 |
| (O) | | 1.068 | 1.014 | 1.131 | .0180712 |
| Szyszkowicz M | 2008 | 1.085 | 1.010 | 1.166 | ***.025766*** |
| Malig BJ | 2013 | 1.020 | 1.010 | 1.030 | ***.0001*** |
| Evans KA | 2013 | 0.821 | 0.418 | 1.403 | .5339787 |
| Ito K | 2007 | 1.060 | 1.052 | 1.072 | ***.0001*** |
| Chardon B | 2007 | 1.044 | 0.999 | 1.104 | .0909348 |
| Lin M | 2002 | 1.011 | 0.925 | 1.065 | .7736097 |
| Silverman RA | 2010 | 1.075 | 1.050 | 1.100 | ***.0001*** |
| Barnett AG (0−4y) | 2005 | 1.045 | 1.018 | 1.071 | ***.000713*** |
| (5−14y) | | 1.034 | 0.992 | 1.076 | .1067006 |
| Iskandar A | 2012 | 1.188 | 1.083 | 1.271 | ***.0001*** |
| Santus P | 2012 | 0.992 | 0.967 | 1.017 | .5433122 |
| Strickland MJ | 2010 | 1.022 | 1.002 | 1.042 | ***.029066*** |
| Andersen ZJ | 2008 | 1.300 | 1.000 | 1.640 | ***.037304*** |
| Hua J | 2014 | 1.003 | 1.000 | 1.010 | .2403955 |
| Gleason JA | 2014 | 1.012 | 1.000 | 1.024 | ***.048279*** |
| Raun LH | 2014 | 1.033 | 0.983 | 1.083 | .1901706 |
| Cheng MH (W) | 2014 | 1.069 | 1.034 | 1.103 | ***.0001*** |
| (C) | | 1.017 | 1.000 | 1.046 | .1420481 |

[1]letters/numbers in brackets indicates results for different population subgroups from the same study.
[2]bold, italicized p-value <.05 calculated after Altman and Bland (2011).

Table 6. Summary information about numbers of base papers and significant results used for meta-analysis for each air quality component.

| Air quality component | Number of risk ratios (RRs) used | RRs with p-values >.05 (%) | RR with p-values ≤.05 | RRs with p-values ≤.001 |
|---|---|---|---|---|
| CO | 42 | 29 (69) | 13 | 9 |
| NO2 | 66 | 30 (45) | 36 | 16 |
| O3 | 71 | 40 (56) | 31 | 11 |
| PM2.5 | 37 | 23 (62) | 14 | 8 |
| PM10 | 51 | 28 (55) | 23 | 6 |
| SO2 | 65 | 46 (70) | 19 | 6 |

Table 7. Asthma characteristics.

---

*Proposed risk factors for development of asthma in children & adults*

*1) Induction (sensitization) phase*

Host factors
genetic predisposition
gender
pre-term delivery (e.g., prematurity
causes lung problems)

Environmental stimuli
allergens
respiratory infections (viruses)

*(2) Maintenance (progression) phase*

Environmental stimuli
allergens
respiratory infections (viruses, bacteria)
tobacco smoke
indoor/outdoor air quality
occupational sensitizers

*Proposed triggering/precipitating factors for exacerbations in children & adults*

emotional factors (stress)
exercise-induced
foods, additives
hormonal factors (premenstrual worsening)
indoor/outdoor allergens
indoor/outdoor air quality
indoor irritants (household sprays, paint fumes)
medications (aspirin, nonsteroidal anti-inflammatory drugs, β-blockers)
occupational agents
substance abuse
tobacco smoke
uncontrolled allergic rhinitis, sinusitis, respiratory viral infections
weather changes (e.g., temperature, humidity)

---

Note: (1) after Lemanske and Busse (2003, 2006); Singh and Busse (2006); Bernstein (2008); Sykes and Johnston (2008); Szefler (2008); Gelfand (2009); Noutsios and Floros (2014); Scherzer and Grayson (2018). (2) All these factors have support in the scientific literature; some are based on associations and some may be directly causative.

List of Figures:

Figure 1. Probability of a research finding in a study being true as a function of disease prevalence rate for various levels of bias (u) (Note: for α=0.05 and power=80%; NPV=negative predictive value – probability that a negative relationship (i.e., risk factor does not cause disease) is true; PPV=positive predictive value – probability that a positive relationship (i.e., risk factor causes disease) is true).

Figure 2. P-value plots for meta-analysis of observational datasets representing: (i) petroleum refinery worker−chronic myeloid leukemia risk and (ii) petroleum refinery worker−mesothelioma risk (after Schnatter et al. 2018).

Figure 3. P-value plots for meta-analysis of observational datasets representing: (i) elderly long-term exercise training−mortality & morbidity risk (after de Souto Barreto et al. 2019) and (ii) smoking−squamous cell carcinoma risk (after Lee et al. 2012).

Figure 4. P-value plots for six air quality components of the Zheng et al. (2015) meta-analysis.

Figure 5. Proposed childhood asthma causes and risks (after Noutsiosa and Floros 2014).

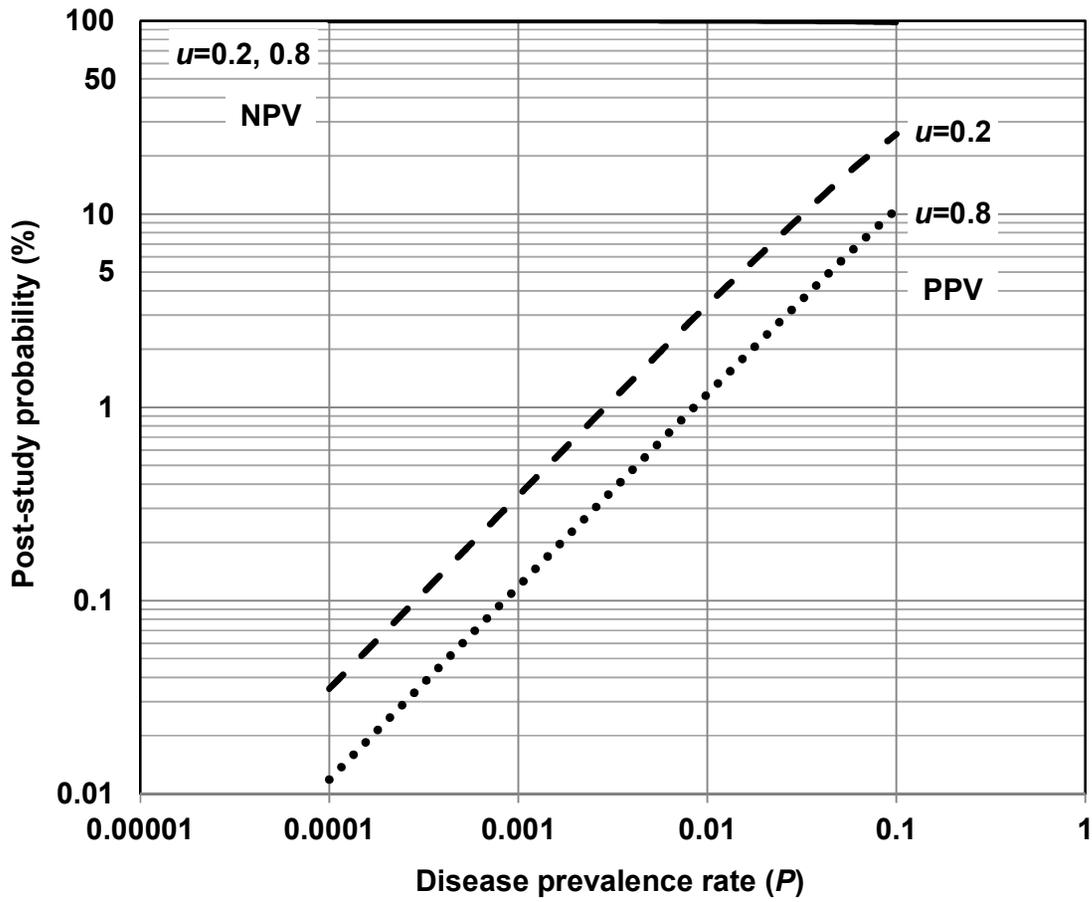

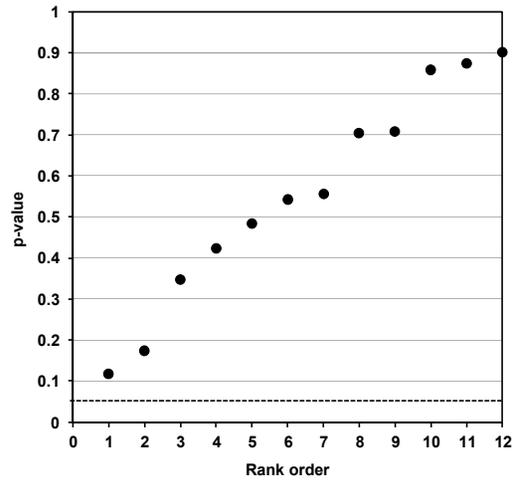 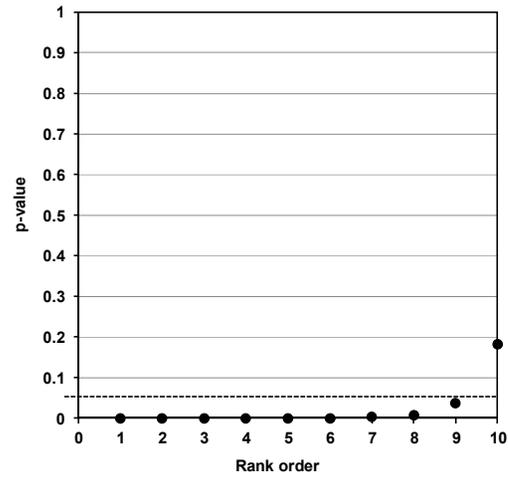

(i) petroleum refinery worker–chronic myeloid leukemia risk (n=12)

(ii) petroleum refinery worker–mesothelioma risk (n=10)

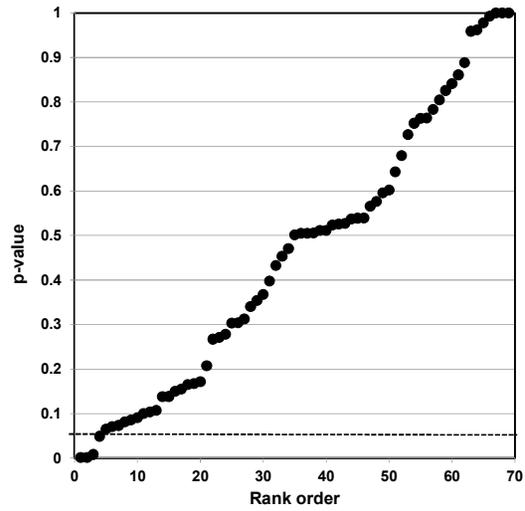 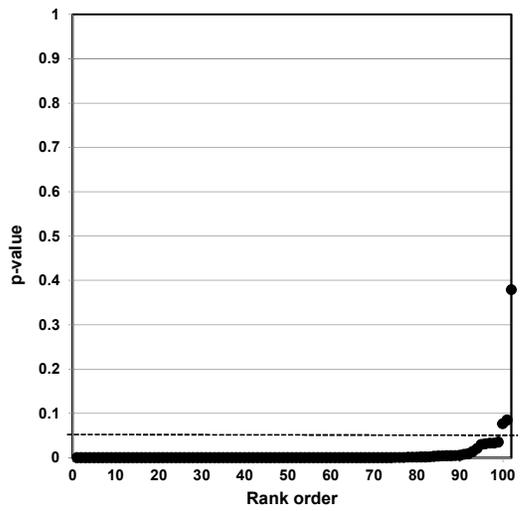

(i) elderly long-term exercise training−mortality & morbidity risk (n=69)

(ii) smoking−squamous cell carcinoma risk (n=102)

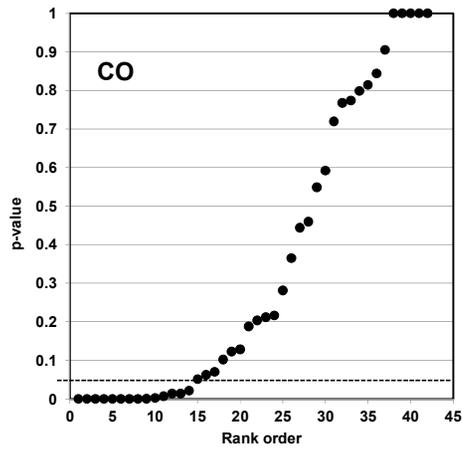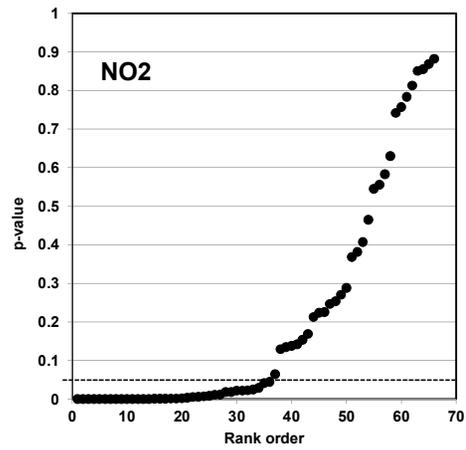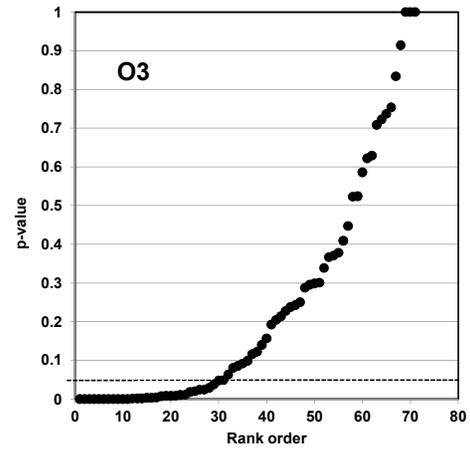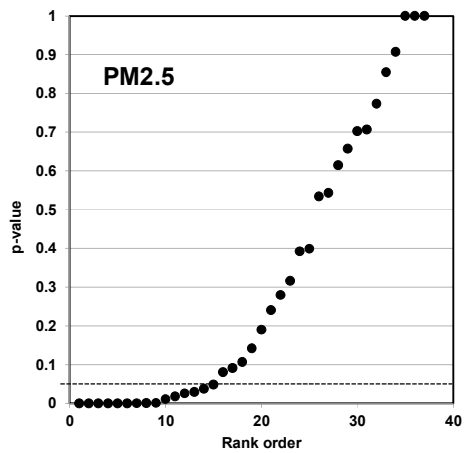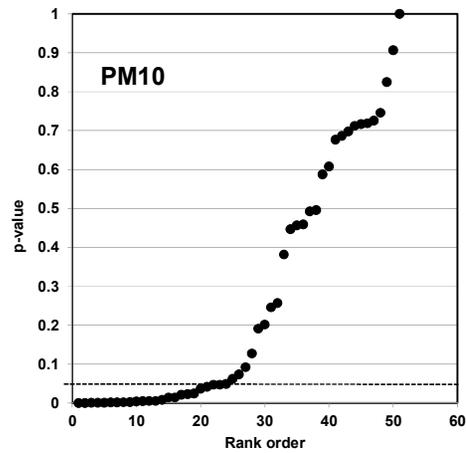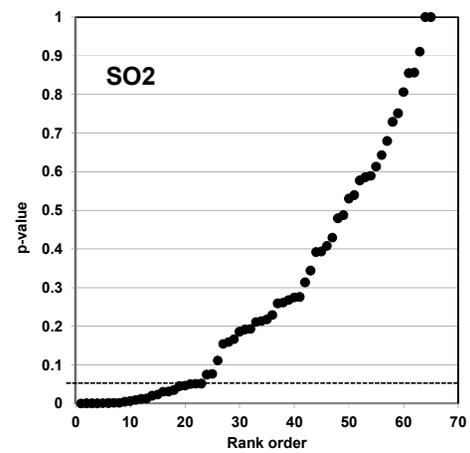

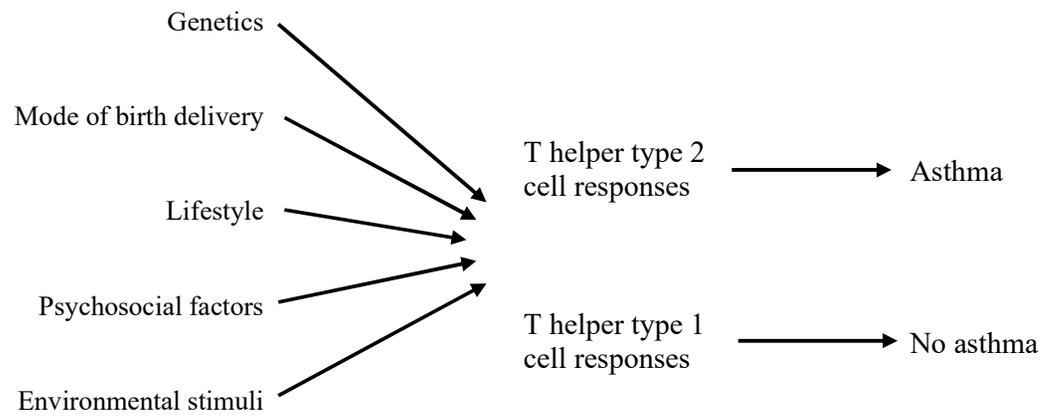

# Evaluation of a meta-analysis of ambient air quality as a risk factor for asthma

# Supplemental Information (SI) files

Warren B. Kindzierski,[a]* S. Stanley Young,[b] Terry G. Meyer[c] and John D. Dunn[d]

[a]School of Public Health, University of Alberta, Edmonton, Alberta, Canada

[b]CGStat, Raleigh, NC, USA

[c]Outcome Based Medicine, Raleigh, NC, USA

[d]401 Rocky Hill Road, Brownwood, TX, USA

*Corresponding author: School of Public Health, University of Alberta, 3-57 South Academic Building, 11405-87 Avenue, Edmonton, Alberta, T6G 1C9 Canada; warrenk@ualberta.ca

SI 1   Estimating the Post-Study Probability of Risk Factor−Chronic Disease Research Findings in Epidemiological Observational Studies

SI 2   List of 87 Base Studies used in the Zheng et al. (2015) Meta-analysis

SI 3   Summary Statistics of Other Air Quality Components and Calculated p-values

SI 4   Summary Statistics of Datasets from Meta-analysis of Selected Cancers in Petroleum Refinery Workers after Schnatter et al. (2018)

SI 5   Summary Statistics of Datasets from Meta-analysis of Elderly Long-term Exercise Training−Mortality & Morbidity Risk after de Souto Barreto et al. (2019)

SI 6   Summary Statistics of Datasets from Meta-analysis of Smoking−Squamous Cell Carcinoma Risk after Lee et al. (2012)

## Supplemental Information, SI 1

## Estimating the Post-Study Probability of Risk Factor−Chronic Disease Research Findings in Epidemiological Observational Studies

*Prevalence rates for various chronic diseases*

Crude estimates of prevalence ($P$) in United States are provided below for selected chronic diseases of interest.

***Respiratory Diseases***

Asthma among total population – $P$ = 0.079 (2017)[a]

Chronic Obstructive Pulmonary Disease (COPD) among adults ≥18 years old – $P$ = 0.059 (during 2014−2015)[b]

***Heart Disease (includes coronary heart disease, angina or heart attack)***

Heart disease among adults ≥18 years old – $P$ = 0.056 (2018)[c]

***Diabetes***

Diabetes among total population – $P$ = 0.094 (2015)[d]

***Cancers***

Breast cancer among females ≥35 years old – $P$ = 0.037 (2016)[e]

Prostate cancer among males ≥55 years old – $P$ = 0.072 (2016)[f]

Colorectal cancer among total population – $P$ = 0.0040 (2016)[g]

Lung & Bronchus among total population – $P$ = 0.0017 (2017)[h]

*Source:*

a    Centers for Disease Control and Prevention, U.S. Department of Health & Human Services, Atlanta, GA – Data, Statistics, and Surveillance, Asthma Surveillance Data, https://www.cdc.gov/asthma/asthmadata.htm (accessed 10 August 2019).

b    COPD Foundation, Denver, CO – National and State Estimates of COPD Morbidity and Mortality — United States, 2014-201, https://journal.copdfoundation.org/jcopdf/id/1209/National-and-State-Estimates-of-COPD-Morbidity-and-Mortality-United-States-2014-2015 (accessed 10 August 2019).

c    Centers for Disease Control and Prevention, U.S. Department of Health & Human Services, Atlanta, GA – Trends in Cancer and Heart Disease Death Rates Among Adults Aged 45–64: United States, 1999–2017, National Vital Statistics Reports, Vol. 68, No. 5, May 22, 2019, https://www.cdc.gov/nchs/data/nvsr/nvsr68/nvsr68_05-508.pdf (accessed 10 August 2019).

d    Centers for Disease Control and Prevention, U.S. Department of Health & Human Services, Atlanta, GA – National Diabetes Statistics Report, 2017, Estimates of Diabetes and Its Burden in the United States, https://www.cdc.gov/diabetes/pdfs/data/statistics/national-diabetes-statistics-report.pdf (accessed 10 August 2019).

e    Estimate of number of females living with breast cancer in 2016: National Cancer Institute, Surveillance, Epidemiology, and End Results (SEER) Program, Bethesda, MD – Cancer Stat Facts: Female Breast Cancer, https://seer.cancer.gov/statfacts/html/breast.html (accessed 16 October 2019) (Note, of females living with breast cancer it was assumed that 4% were <35 yr old and 96% ≥35 yr old) & Estimate of number of females living in 2016: United States of America 2016 population, https://www.populationpyramid.net/united-states-of-america/2016/ (accessed 16 October 2019).

f    Estimate of number of males living with prostate cancer in 2016: National Cancer Institute, SEER Program, Bethesda, MD – Cancer Stat Facts: Prostate Cancer, https://seer.cancer.gov/statfacts/html/prost.html (accessed 16 October 2019) (Note, of males living with prostate cancer it was assumed that 2% were <55 yr old and 98% ≥55 yr old) &

Estimate of number of males living in 2016: United States of America 2016 population, https://www.populationpyramid.net/united-states-of-america/2016/ (accessed 16 October 2019).

g    Estimate of number of people living with colorectal cancer in 2016: National Cancer Institute, SEER Program, Bethesda, MD – Cancer Stat Facts: Colorectal Cancer, https://seer.cancer.gov/statfacts/html/colorect.html (accessed 16 October 2019) & Estimate of number of people living in 2016: United States of America 2016 population, https://www.populationpyramid.net/united-states-of-america/2016/ (accessed 16 October 2019).

h    Estimate of number of people living with lung and bronchus cancer in 2016: National Cancer Institute, SEER Program, Bethesda, MD – Cancer Stat Facts: Lung and Bronchus Cancer, https://seer.cancer.gov/statfacts/html/lungb.html (accessed 16 October 2019) & Estimate of number of people living in 2016: United States of America 2016 population, https://www.populationpyramid.net/united-states-of-america/2016/ (accessed 16 October 2019).

*Post-study probabilities of a research finding being true*

Our interest is in estimating the post-study probability of an outcome being true for a single observational study of risk factor−chronic disease relationships. An outcome may either be one that is a positive (+ve) association, i.e. one that is statistically significant with p-value ≤.05; or a negative (−ve) association, i.e. one that is not statistically significant with p-value >.05. Using a Bayesian argument, the former outcome may either be a true or false positive association and the latter outcome may either be a false or true negative association.

*Disease prevalence (P)*

First we provide a formal definition of prevalence ($P$):

$P =$ The number of persons with a disease divided by the total population at risk at a particular point in time (Wassertheil-Smoller 1995).[a]

We break this down further as:

$d^+ =$ Those in the population under study with the disease.

$d^- =$ Those in the population under study without the disease.

$d^+ + d^- =$ Total population at risk under study.

$\therefore P = d^+/(d^+ + d^-)$  (1)

Odds of a particular outcome, in this case odds in favor of $d^+$ given $d^+$ and $d^-$, is:[a]

$\text{Odds}(d^+) = d^+/d^-$  (2)

Probability (P) of outcome $d^+$ is related to odds:[a]

$P(d^+) = \text{Odds}(d^+)/[\text{Odds}(d^+) + 1]$  (3)

Equation (2) is the same as the Ioannidis (2005)[b] definition of $R$ being the "number of *true relationships* to *no relationships* among those tested in the field" (i.e., $\text{Odds}(d^+) \equiv R$). Also, equation (3) is the same as the Ioannidis (2005)[b] definition of "pre-study probability of a relationship being true" (i.e., $P(d^+) = \text{Odds}(d^+)/[\text{Odds}(d^+) + 1] \equiv R/(R + 1)$).

If we substitute equation (2) into equation (3) and rearrange, we get $P(d^+) = d^+/(d^+ + d^-)$, which is the same as the prevalence rate of a disease ($P$). Thus the "pre-study probability of a relationship being true" can be represented by the prevalence rate of a disease ($P$) for our purposes. As our interest is in estimating the post-study probability of an outcome being true as a function of $P$, we conservatively assume that all those in the population with the disease (i.e., $d^+$) being studied are due to the specific risk factor we are examining.

*2 x 2 Table of true relationships and study findings in absence of bias*

Now let us consider a 2 x 2 table of measures of association between 'true relationships' and 'study findings' and in the absence of bias for a single observational study of risk factor−chronic disease relationships (refer to *Illustration* 1). Hrudey and Leiss (2003)[c] and Baduashvili et al. (2020)[d] are used to derive our 2 x 2 table.

|  |  | True Relationship of a Risk Factor | | |
|---|---|---|---|---|
|  |  | Causes disease | Does not causes disease | Total |
| **Study Finding** | Causes disease | **True Positive (TP)** with rate = Se (Sensitivity) = $1 - \beta$ | **False Positive (FP)** with rate = $\alpha$ | = TP + FP |
|  | Does not cause disease | **False Negative (FN)** with rate = $\beta$ | **True Negative (TN)** with rate = Sp (Specificity) = $1 - \alpha$ | = FN + TN |
|  | Total | = TP + FN | = FP + TN | = $c$ |

***Illustration* 1. 2 x 2 table of true relationships and study findings given *c* relationships probed/tested** (after Hrudey and Leiss, 2003[c] and Baduashvili et al. 2020[d]).

For $c$ risk factor−disease relationships probed/tested in a single study, we can define the following from *Illustration* 1:

Number of true +ve associations, $N(TP) = c$ x Se (sensitivity) x $P = c(1 - \beta)P$     (4)

Number of false −ve associations, $N(FN) = c\beta P$     (5)

Number of true −ve associations, $N(TN) = c$ x Sp (specificity) x $(1 - P) = c(1 - \alpha)(1 - P)$     (6)

Number of false +ve (FP) associations, $N(FP) = c\alpha(1 - P)$     (7)

where $\alpha$ and $\beta$ are the Type I and Type II error rates, respectively.

The total number of positive associations, $N[(TP) + (FN)] =$
$c(1 - \beta)P + c\beta P = cP$

The total number of negative associations, N[(FP) + (TN)] =
$c\alpha(1 - P) + c(1 - \alpha)(1 - P) = c(1 - P)$

As a check of the bottom row, the total number of positive and negative associations =
N[(TP) + (FN)] + N[(FP) + (TN)] = $cP + c(1 - P) = c$

Now we do a check of the far-right column:
N[(TP) + (FP)] = $c(1 - \beta)P + c\alpha(1 - P)$
N[(FN) + (TN)] = $c\beta P + c(1 - \alpha)(1 - P)$
∴ N[(TP) + (FP)] + N[(FN) + (TN)] = $c(1 - \beta)P + c\alpha(1 - P) + c\beta P + c(1 - \alpha)(1 - P)$;
rearranging… = $\{c(1 - \beta)P + c\beta P\} + \{c\alpha(1 - P) + c(1 - \alpha)(1 - P)\} = cP + c(1 - P) = c$

*2 x 2 Table of true relationships and study findings in presence of bias*

Now we consider the role of bias ($u$) in an observational study of risk factor−chronic disease relationships. Here we use the definition of $u$ after Ioannidis (2005)[b] – *the proportion of probed analyses* [relationships] *that would not have been "research findings," but nevertheless end up presented and reported as such, because of bias*. "Research findings" imply nominally significant outcomes (i.e., associations with p-value ≤.05). Bias will transfer a portion of non-significant or null outcomes (associations with p-value >.05) to nominally significant outcomes (associations with p-value ≤.05).

First, we deal with positive associations – FNs & TPs. Because of bias, some outcomes initially showing up as non-significant FN associations (i.e., with p-value >.05) will now end up being reported as significant TP associations (i.e., with p-value ≤.05) proportional to $u$:
∴ N(FN) in the presence of bias now becomes = $c\beta P(1 - u)$     (8)
∴ N(TP) in the presence of bias now becomes = $c(1 - \beta)P + uc\beta P$     (9)

The total number of positive associations, N[(TP) + (FN)] =
$c(1 - \beta)P + uc\beta P + c\beta P(1 - u) = cP$

Now we deal with negative associations – TNs & FPs. Because of bias, some outcomes initially showing up as non-significant TN associations (i.e., with p-value >.05) will now end up being reported as significant FP associations (i.e., with p-value ≤.05) proportional to $u$:

∴ N(TN) in the presence of bias now becomes = $c(1 - \alpha)(1 - P)(1 - u)$     (10)

∴ N(FP) in the presence of bias now becomes = $c\alpha(1 - P) + uc(1 - \alpha)(1 - P)$     (11)

The total number of negative associations, N[(FP) + (TN)] =

$c\alpha(1 - P) + uc(1 - \alpha)(1 - P) + c(1 - \alpha)(1 - P)(1 - u)$; rearranging =

$\{c\alpha(1 - P) + c(1 - \alpha)(1 - P)\} + \{[uc(1 - \alpha)(1 - P)] - [uc(1 - \alpha)(1 - P)]\} = c(1 - P)$

As a check of the bottom row, the total number of positive and negative associations =

N[(TP) + (FN)] + N[(FP) + (TN)] = $cP + c(1 - P) = c$

Now we do a check of the far-right column:

N[(TP) + (FP)] = $c(1 - \beta)P + uc\beta P + c\alpha(1 - P) + uc(1 - \alpha)(1 - P)$

N[(FN) + (TN)] = $c\beta P(1 - u) + c(1 - \alpha)(1 - P)(1 - u)$

∴ N[(TP) + (FP)] + N[(FN) + (TN)] = $c(1 - \beta)P + uc\beta P + c\alpha(1 - P) + uc(1 - \alpha)(1 - P) +$

$c\beta P(1 - u) + c(1 - \alpha)(1 - P)(1 - u)$; rearranging =

$\{c\alpha(1 - P) + c(1 - \alpha)(1 - P)\} + cP + \{[uc(1 - \alpha)(1 - P)] - [uc(1 - \alpha)(1 - P)]\} + \{uc\beta P - uc\beta P\} + \{c\beta P - c\beta P\} = c(1 - P) + cP = c$

*Post-study probability of an outcome being true*

The positive predictive value (PPV)$^c$ is used to represent the post-study probability of a positive outcome being true in the presence of bias:

PPV = N(TP)/N[(TP) + (FP)]     (12)

∴ PPV = $[(1 - \beta)P + u\beta P] / [(1 - \beta)P + u\beta P + \alpha(1 - P) + u(1 - \alpha)(1 - P)]$     (13)

The negative predictive value (NPV)$^c$ is used to represent the post-study probability of a negative (i.e., null) outcome being true in the presence of bias:

NPV = N(TN)/N[(FN) + (TN)]     (12)

∴ NPV = $[(1 - \alpha)(1 - P)(1 - u)] / [\beta P(1 - u) + (1 - \alpha)(1 - P)(1 - u)]$     (13)

# Supplemental Information, SI 2

## List of 87 Base Studies used in the Zheng et al. 2015 Meta-analysis

(Note: 'number' preceding reference is the citation number used by Zheng et al.)

# 17 Base Studies Randomly Selected for Evaluation of Analysis Search Space

(Note: 'number' preceding reference is the citation number used by Zheng et al.)

Initial 10 base studies selected:

3. Thompson AJ, Shields MD. Acute asthma exacerbations and air pollutants in children living in Belfast, Northern Ireland. Arch Eviron Heal 2001; 56: 234–241.

18. Tsai SS, et al. Air pollution and hospital admissions for asthma in a tropical city: Kaohsiung, Taiwan. Inh Toxicol 2006; 18: 549–554.

23. Laurent O, et al. Air pollution, asthma attacks, and socioeconomic deprivation: a small-area case-crossover study. Am J Epidemiol 2008; 168: 58–65.

44. Lin M, et al. The Influence of ambient coarse particulate matter on asthma hospitalization in children: case-crossover and time-series analyses. Environ Health Persp 2002; 110: 575–581.

47. Chakraborty P, et al. Effect of airborne Alternariaconidia, ozone exposure, $PM_{10}$ and weather on emergency visits for asthma in school-age children in Kolkata city, India. Aerobiologia 2014; 30: 137–148.

48. Abe T, et al. The relationship of short-term air pollution and weather to ED visits for asthma in Japan. Am J Emerg Med 2007; 27: 153–159.

69. Evans KA, et al. Increased ultrafine particles and carbon monoxide concentrations are associated with asthma exacerbation among urban children. Environ Res 2014, 129: 11–19.

70. Sheppard L, et al. Effects of ambient air pollution on nonelderly asthma hospital admissions in Seattle, Washington, 1987–1994. Epidemiology 1999; 10: 23–30.

80. Gouveia N, Fletcher T. Respiratory diseases in children and outdoor air pollution in San Paulo, Brazil: a time series analysis. Occup Environ Med 2000; 57: 477–483.

83. Hua J, et al. Acute effects of black carbon and $PM_{2.5}$ on children asthma admissions: A time-series study in a Chinese city. Sci Tot Envir 2014; 481: 433–438.

Additional 7 base studies selected:

8. Andersen ZJ, et al. Size distribution and total number concentration of ultrafine and accumulation mode particles and hospital admissions in children and the elderly in Copenhagen, Denmark. Occup Environ Med 2008; 65: 458–466.

15. Chardon B, et al. Air pollution and doctors' house calls for respiratory diseases in the Greater Paris area (2000–3). Occup Environ Med 2007; 64: 320–324.

18. Tsai SS, et al. Air pollution and hospital admissions for asthma in a tropical city: Kaohsiung, Taiwan. Inh Toxicol 2006; 18: 549–554.

31. Tenias JM, et al. Association between hospital emergency visits for asthma and air pollution in Valencia, Spain. Occup Environ Med 1998; 55: 541–547.

35. Magas OK, et al. Ambient air pollution and daily pediatric hospitalizations for asthma. Env Sci Pollut Res 2007; 14: 19–23.

57. Lavigne E, et al. Air pollution and emergency department visits for asthma in Windsor, Canada. Can J Public Heal 2012; 103: 4–8.

64. Mar T, et al. Associations between asthma emergency visits and particulate matter sources, including diesel emissions from stationary generators in Tacoma, Washington. Inhal Toxicol 2010; 22: 445–448.

74. Santus P, et al. How air pollution influences clinical management of respiratory diseases. A case-crossover study in Milan. Respir Res 2012; 13: 95.

# Details of the 87 base studies included in the Zheng et al. 2015 meta-analysis

Notes: Data presented in the tables below obtained from Zheng et al. (2015).
Study: '#', 'Name', 'year': Zheng et al. 2015 citation number, first author, year published.
Quality score: 0 (low) to 5 (high quality); ICD: International Classification of Diseases; ICPC2: International Classification of Primary Care 2; GP: GP's house calls;
NAEPP: National Asthma Education and Prevention Program; PC: primary care visits; TS: time-series; CC: case crossover.
Default concentration for CO in mg/m³ and for NO₂, SO₂, PM₁₀ and PM₂.₅ all in μg/m³ is 24-h averaged concentration; and for O₃ in μg/m³ is 8-hr max concentration.

| Study | Location, Period | Quality score | Sample size | Population | Type of study | Pollutants & concentration | | | | | | Measurement quality score (0 or 1 point) | Lag pattern (single day/ mean days) | Adjustments (long-term trend, seasonality, temperature, humidity, pressure, day for the week, holiday and influenza epidemics) |
|---|---|---|---|---|---|---|---|---|---|---|---|---|---|---|
| | | | | | | CO | PM₁₀ | PM₂.₅ | SO₂ | NO₂ | O₃ | | | |
| **Base studies with emergency room visit indices:** | | | | | | | | | | | | | | |
| 3, Thompson AJ, 2001 | Belfast, Northern Ireland, 3 yr | 5 | - | Children | TS | 0.65 | 28.4 | - | 47.1 | 43.6 | 38.4 | 1 | Both | long-term trend, seasonality, temperature, humidity, day of week, public holiday |
| 14, Sunyer J, 1997 | Barcelona, Helsinki, Paris, London, 6 yr | 5 | 75190 | Children + adults | TS | | | | 27.8 | 49.8 | 43.8 | 1 | Both | long-term trend, seasonality, temperature, humidity, day-of-week, influenza epidemics |
| 15, Chardon B, 2007 | Paris, 3 yr | 5/4[b] | 8027 | General | TS | | 23.0 | 14.7 | | 44.4 | | 1/0[b] | Cumulative | long-term trends, seasonality, temperature, humidity, day-of-week, influenza epidemic, pollen |
| 16, Halonen JI, 2008 | Helsinki, Finland, 6 yr | 5 | 4807 | General | TS | 0.5[a] | | 9.5 | | 28.2 | | 1 | Single | long-term trends, seasonality, temperature, humidity, day-of-week, public holiday, influenza epidemics |
| 30, Hajat S, 1999 | London,UK, 3 yr | 5 | 38653 | General | TS | 1.0 | 28.5 | | 21.2 | 69.0 | 37.5 | 1 | Both | long term trend, seasonality, temperature, humidity, day-of-week, influenza epidemic |
| 41, Galan I, 2003 | Madrid, Spain, 3 yr | 5 | 4827 | General | TS | | 32.1 | | 23.6 | 67.1 | 45.8 | 1 | Single | long-term trend, seasonality, temperature, humidity, pressure, day-of-week, public holiday, influenza epidemics |

| Study | Location, duration |  |  | Population | Design | Col3 | Col4 | Col5 | Col6 | Col7 | Col8 | Col9 | Exposure | Confounders |
|---|---|---|---|---|---|---|---|---|---|---|---|---|---|---|
| 51, Atkinson RW, 2001 | Barcelona, Birmingham, London, Milan, the Netherlands, Paris, Rome, and Stockholm, 4 yr | 5 | - | General | TS | | 29.3 | | | | | 1 | Single | long-term trend, seasonality, temperature, humidity, day for the week, holiday and influenza epidemics |
| 55, Medina S, 1997 | Paris, France, 4 yr | 5 | - | General | TS | | | | 19.0 | 56.0 | 34.0 | 1 | Both | long-term trend, seasonality, temperature, humidity, day-of-week, influenza epidemics |
| 58, Stieb DM, 2009 | Canada (Montreal, Ottawa, Edmonton, Saint John, Halifax, Toronto, Vancouver), 10 yr | 5/4[b] | 83563 | General | TS | 0.8 | 8.3 | 20.6 | 14.6 | 37.6 | 40.0 | 0/1[a] | Single | long-term trend, seasonality, temperature, humidity, day-of-week, holiday |
| 60, Wilson AM, 2005 | Portland, Maine, Manchester, New Hampshire, 2 yr | 5 | 7300 | General | TS | | | | 39.4[b] | | 41.0 | 1/0(ozone) | Single | long-term trend, seasonality, temperature, humidity, day-of-week, influenza epidemic |
| 75, Sunyer J, 2003 | Birmingham, London, Milan, the Netherlands, Paris, Rome, Stockholm, 2-8 yr | 5 | - | Children + adults | TS | | | | 17.6 | | | 1 | Single | long-term trend, seasonality, temperature, humidity, day-of-week, holiday, influenza epidemic |
| 27, Szyszkowicz M, 2008 | Edmonton, Canada, 10 yr | 4 | 62563 | General | TS | 0.9 | 22.6 | 8.5 | | 45.0 | 39.9 | 1 | Single | long term trend, temperature, seasonality humidity, day-of-week |
| 29, Norris G, 1999 | Seattle, US, 27 mon | 4 | 1458 | Children | TS | | 21.7 | | 17.1 | 41.5 | | 1 | Single | long term trend, seasonality temperature, humidity, day-of-week |
| 31, Tenias JM, 1998 | Valencia, Spain, 3 yr | 4 | 734 | Adults + elderly | TS | | | | 26.6 | 57.7 | 62.8 | 1 | Single | long-term trend, seasonality, temp., humidity, day-of-week, public holiday, influenza epidemics |
| 32, Stieb DM, 1996 | Saint John, New Brunswick, Canada, 8 yr | 3 | 1163 | General | TS | | | | | | 89.1 | 1 | Single | long-term trend, seasonality, temperature, humidity, day-of-week |
| 43, Mohr LB, 2008 | St. Louis, US, 2 yr | 3 | 12836 | Children | TS | | | - | | - | - | 1 | Single | long-term trend, seasonality, temperature |

| Study | Location, Duration | Lag | N | Population | Design |  |  |  |  |  |  |  | Pollutant | Confounders |
|---|---|---|---|---|---|---|---|---|---|---|---|---|---|---|
| 46, Castellsague J, 1995 | Barcelona, Spain, 5 yr | 3 | 6019 | Adults+ elderly | TS | | | | 45.5 | 58.0 | 70.5 | 1 | Cumulative | long-term trend, seasonality, temperature, humidity, day-of-week, influenza epidemic |
| 47, Chakraborty P, 2013 | Kolkata city, India, 2 yr | 4 | 2703 | Children | TS | | 3.8 | | | | 1.3[c] | 1 | - | long term trend, seasonality, temperature, humidity |
| 56 Jaffe DH, 2003 | Cincinnati, Cleveland, and Columbus, 5 yr | 4 | 4416 | Children+ adults | TS | | 49.5 | | 28.8 | 10.1 | 35.8 | 1 | Single | long-term trend, seasonality, temperature, humidity, day-of-week |
| 64, Mar TF, 2010 | Tacoma, Washington, 3.5 yr | 3 | 10091 | General | TS | 1.2 | | 12.3 | | | | 1 | Single | long-term trend, seasonality, temperature, humidity |
| 66, Cirera L, 2012 | Cartagena, Spain, 4 yr | 4 | 1617 | General | TS | | | | 32 | 51 | 81 | 1 | Single | long-term trend, seasonality, temperature, humidity, day-of-week, public holiday, influenza epidemic |
| 67, Cassino C, 1999 | New York, US, 3.5 yr | 4 | 285 | Adults | TS | 1.2 | | | 29.4 | 92.4 | 37.5[c] | 1 | Single | long-term trend, seasonality, temperature, humidity, day-of-week |
| 76, Babin S, 2008 | Washington DC, US, 11 yr | 4 | 61218 | General | TS | | - | - | | | - | 1 | Single | long-term trend, seasonality, temperature, humidity, day-of-week |
| 78, Mar TF, 2009 | Seattle, Washington, US, 4 yr | 3 | 3217 | Children + adults | TS | | | | | | 84.0 | 0 | Single | long-term trend, seasonality, temperature, humidity, day-of-week |
| 79, Hernandez-Cadena L, 2000 | Ciudad Juárez, Chihuahua, Mexico, 1 yr | 4 | 2459 | General | TS | | 34.46 | | | | 110.6 | 1 | Both | long-term trend, seasonality, temperature, humidity, day-of-week |
| 82, Ito K, 2007 | New York, US, 4 yr | 3 | | General | TS | 1.6[a] | 15.7 | | 22.3 | 31.1 | 65.1 | 1 | Both | long-term trend, seasonality, temperature, humidity, day-of-week |
| 84, Cadelis G, 2014 | Guadeloupe, France, 1 yr | 3 | 836 | Children | TS | | 19.2 | | | | | 1 | Both | long-term trend, seasonality, temperature |
| 42, Jazbec A, 1999 | Zagreb, Croatia, 1.5 yr | 0 | 1372 | Children +adults | TS | | | | | 45.1 | | 0 | Cumulative | long-term trend, seasonality, |
| 48, Abe T, 2007 | Tokyo, Japan, 1 yr | 2 | 6447 | General | TS | 1.4 | | | 15.1 | | | 1 | Single | seasonality, temperature, humidity |
| 61, Chimonas MAR, 2007 | Anchorage, Alaska, 3.5 yr | 2 | 11037 | Children | TS | | 27.6 | 6.1 | | | | 0 | Single | long-term trend, seasonality, temperature |
| 4, Strickland MJ, 2010 | Atlanta, US, 11 yr | 5 | 91386 | Children | CC | 1.1[b] | 23.8 | 16.4 | 30.9[b] | 47.8[b] | 97.3 | 1 | Both | long-term trend, seasonality, temperature, humidity, day of week and influenza epidemics |

| Study | Location, duration | Col1 | N | Population | Design |  |  |  |  |  |  |  | Lag | Confounders |
|---|---|---|---|---|---|---|---|---|---|---|---|---|---|---|
| 11, Paulu C, 2008 | Maine, US, 4 yr | 5 | 8020 | General | CC | | | 8.5 | | | 83.6 | 1 | Both | long-term trend, seasonality, temperature, humidity, day of week, holiday |
| 54, Jalaludin BB, 2008 | Sydney, Australia, 5 yr | 5 | 317724 | Children | CC | 1.0 | 16.8 | 9.4 | 3.1 | 47.6 | 67.7 | 1 | Both | long-term trend, seasonality, temperature, humidity, day-of-week, public holiday |
| 57, Lavigne E, 2010 | Windsor, Canada, 7 yr | 5 | 3728 | General | CC | 0.4 | | 7.3 | 5.3 | 17.4 | 41.5 | 1 | Single | long-term trend, seasonality, temperature, humidity, day-of-week, influenza epidemic |
| 22, Malig BJ, 2013 | California, 4 yr | 4 | 74978 | General | CC | | 35.0 | 12.1 | | | | 1 | Single | long-term trends, seasonality, temperature, humidity, day of week |
| 49, Boutin-Forzano S, 2004 | Marseille, France, 1 yr | 3 | 549 | Children + adults | CC | | | | 22.5 | 34.9 | 50.1 | 1 | Single | long-term trend, seasonality, temperature, humidity |
| 72, Yamazaki S, 2013 | Himeji, Japan, 2 yr | 3 | 956 | Children | CC | | 34.3 | 21.2 | | 22.9 | 55.3 | 1 | Single | long-term trend, seasonality, temperature, humidity, day-of-week |
| 74, Santus P, 2012 | Milan, Italy, 2 yr | 3 | 3569 | General | CC | 1.5 | 47.1 | 32.8 | 4.13 | 102.6 | 74.3 | 1 | Both | long-term trend, seasonality, temperature, humidity, day-of-week |
| 77, Yamazaki S, 2009 | Tokyo, Japan, 1 yr | 4 | 403 | Children+ adults | CC | | | 19.1 | | 45.3 | 60.2 | 1 | Cumulative | long-term trend, seasonality, temperature, humidity, day-of-week, holiday |
| 6, Mehta AJ, 2012 | Switzerland, 2 yr | 1 | 147 | adults | CC | | | | | 33.9 | | 1 | Single | long-term trend, influenza epidemics |
| 23, Laurent O, 2008 | France, 5 yr | 2 | 4677 | General | CC | | 22.6 | | 8.9 | 36.0 | 57.7 | 1 | Both | temperature, pressure, humidity, influenza epidemic and pollen count |
| 50, Pereira G, 2010 | Perth, Australia, 5 yr | 2 | 603 | Children | CC | 0.3 | | | | 12.9 | | 1 | Single | not declared |
| 69, Evans KA, 2013 | New York, US, 3 yr | 0 | 71 | Children | CC | 0.5 | | 8.6 | 15.4 | | 55.9 | 0 | Both | temperature, humidity |
| 87, Gleason JA, 2014 | New Jersey, US, 4 yr | 5 | 21,854 | Children | CC | | | - | | | - | 1 | Both | long-term trend, seasonality, temperature, humidity, day-of-week, holiday, viral upper respiratory infections |
| 88, Sacks JD, 2014 | North Carolina, US, 3 yr | 5 | 121,621 | General | CC | | | | | | 93.4 | 1 | Cumulative | long-term trend, seasonality, temperature, humidity, day-of-week, holiday, viral upper respiratory infections |

| Study | Location, Period | Quality score | Sample size | Population | Type of study | Pollutants and concentration (CO in units of mg/m³; all other pollutants in units of μg/m³) | | | | | | Measurement quality score (0-1 point) | Lag pattern (single day/ mean days) | Adjustments (long-term trend, seasonality, temperature, humidity, pressure, day for the week, holiday and influenza epidemics) |
|---|---|---|---|---|---|---|---|---|---|---|---|---|---|---|
| | | | | | | CO | PM$_{10}$ | PM$_{2.5}$ | SO$_2$ | NO$_2$ | O$_3$ | | | |
| 89, Raun LH, 2014 | Houston, Texas, US, 7 yr | 3 | 11,754 | General | CC | 0.3 | | 10.7 | 4.3 | 21.6 | 76.7 | 0 | Both | long-term trend, seasonality, temperature, humidity, day-of-week, holiday, viral upper respiratory infections |
| **Base studies with hospital admission indices:** | | | | | | | | | | | | | | |
| 7, Samoli E, 2010 | metropolitan area of Athens, 3 yr | 5 | 3601 | Children | TS | | 43.9 | | 16.8 | 84.8 | 70.9 | 1 | Single | long-term trends, seasonality, temperature, humidity, day of week, public holidays and influenza epidemics |
| 19, Wong TW, 1999 | Hong Kong, 1 yr | 5 | - | General | TS | | 45.0 | | 17.1 | 51.4 | 24.2 | 1 | Both | long-term trends, seasonality, temperature, humidity, days-of-week, public holiday |
| 20, Fusco D, 2001 | Rome, Italy, 2 yr | 5 | 4635 | General | TS | 3.6 | | | 9.1 | 86.7 | 27.0 | 1 | Single | long-term trends, seasonality, temperature, humidity, day- of-week, public holiday, influenza |
| 21, Morgan G, 1998 | Sydney, Australia, 5 yr | 5 | - | General | TS | | 19.2 | | | 30.8 | 53.6 | 1 | Single | long-term trends, seasonality, temperature, humidity, day-of-week, holiday |
| 24, Anderson HR, 1998 | London, UK, 5 yr | 5 | 63039 | General | TS | | | | 32 | 76.4 | 33.2 | 1 | Both | Long-term trends, temperature, humidity, seasonality, days of the week, public holidays and influenza epidemics |
| 28, Lee SL, 2006 | Hong Kong, 5 yr | 5 | 26663 | Children | TS | | 56.1 | 45.3 | 17.7 | 64.7 | 28.6 | 1 | Single | Long-term trend, seasonality, temperature, humidity, day-of-week, public holiday, influenza epidemic |

| Ref, Author, Year | Location, Duration | | N | Population | Model | | | | | | | Lag | Exposure | Confounders |
|---|---|---|---|---|---|---|---|---|---|---|---|---|---|---|
| 34, Petroeschevsky A, 2001 | Brisbane, Australia, 8 yr | 5 | 13246 | Children + adults | TS | | | | 11.7 | 28.5 | 40.7 | 1 | Both | long term trend, seasonality, temperature, humidity, day-of-week, influenza epidemic |
| Ko FWS38 2007 | Hong Kong, China, 6yr | 5 | 69176 | General | TS | | 52.5 | 36.4 | 18.8 | 53.2 | 43.4 | 1 | Both | long term trend, seasonality, temperature, humidity, day-of-week, public holiday |
| 39, Krmpotic D, 2011 | Zagreb, Croatia, 3 yr | 5 | 808 | Adults | TS | 0.8 | 39.0 | | | 30.3 | | 1 | Single | long term trend, seasonality, temperature, humidity, day-of-week, influenza epidemic |
| 59, Romero-Placeres M, 2004 | Habana, Cuba, 2 yr | 5 | 44 029 | General | TS | | 59.2 | | 21.1 | | | 1 | Single | long-term trend, seasonality, temperature, humidity, day-of-week, public holiday |
| 71, Morgan G, 2010 | Sydney, Australia, 8.5 yr | 5 | 65448 | Children + adults | TS | | 62.0 | | | | | 1 | Single | long-term trend, seasonality, temperature, humidity, day of week, influenza epidemic |
| 80, Fletcher T, 2000 | San Paulo, Brazil, 2 yr | 4/5 | - | Children | TS | 5.8 | 64.9 | | 18.3 | 174.3 | 63.4[c] | 1 | Single | long-term trend, seasonality, temperature, humidity, day of week, holiday |
| 5, Son JY, 2013 | Seoul, Busan, Incheon, Daegu, Daejeon, Gwangju, Ulsan, Korea, 6 yr | 4 | - | General | TS | 1.0[a] | 52.4 | | 15.7 | 48.0 | 71.7 | 1 | Both | long-term trend, seasonality, temperature, humidity, pressure, day-of-week |
| 8, Andersen ZJ, 2008 | Copenhagen, Denmark, 3yr | 4 | - | Children | TS | 0.36 | 24.0 | 10.0 | | 22.6 | 51.4 | 0 | Cumulative | long-term trends, seasonality, temperature, humidity, day-of-week, public holidays, influenza epidemics |
| 10, Kim SY, 2012 | Denver, US, 4 yr | 4 | 10,590 | General | TS | | | 8.0 | | | | 1 | Single | long-term trend, seasonality, temperature, humidity, day-of-week |
| 12, Lee JT, 2002 | Seoul, Korea, 2 yr | 4 | 6436 | General | TS | 2.3 | 64.0 | | 22 | 64.7 | 77.1 | 1 | Cumulative | long-term trend, seasonality, temperature, and humidity, day-of-week |
| 13, Delfino RJ, 1994 | Montreal, Canada, 4 yr | 3 | 10385 | General | TS | | 29.5 | | | | | 0 | Single | long-term trend, seasonality, temperature, humidity, day-of-week |

| Ref | Location, period | Lag | N | Population | Design | [col7] | [col8] | [col9] | [col10] | [col11] | [col12] | Dummy | Pollutant | Confounders |
|---|---|---|---|---|---|---|---|---|---|---|---|---|---|---|
| 25, Ye F, 2001 | Tokyo, Japan, 15 yr | 3 | 2200 | Elderly | TS | | 46.0 | | | | | 1 | Single | long-term trends, seasonality, temperature |
| 36, Lee JT, 2006 | Seoul, Korea, 1 yr | 4 | 2952 | Children | TS | 8.0[a] | 135.2 | | 25.0 | 156.2 | 63.9 | 1 | Single | Long-term trend, seasonality, temperature, humidity, day-of-week, |
| 40, Lin M, 2004 | Vancouver, BC, Canada, 12 yr | 4 | 3754 | Children | TS | 1.2 | | | 13.6 | 38.4 | 60.0[c] | 1 | Both | long term trend, seasonality, temperature, humidity, day-of-week |
| 45, Schouten JP, 1996 | Amsterdam, The Netherlands, 12 yr | 4 | - | General | TS | | | | 28.0 | 50.0 | 69.0 | 0 | Both | long-term trend, seasonality, temperature, humidity, day-of-week, public holiday, influenza epidemics |
| 52, Silverman RA, 2010 | New York, US, 7 yr | 4 | 75383 | General | TS | | | 13.0 | | | 87.9 | 1 | Cumulative | long-term trend, seasonality, temperature, humidity, day-of- week |
| 65, Amancio CT, 2012 | San Paulo, Brazil, 2 yr | 4 | 841 | Children | TS | | 25.2 | | 4.6 | | 74.3 | 1 | Single | long-term trend, seasonality, temperature, humidity, day-of-week |
| 70, Sheppard L, 1999 | Seattle, Washington, 8 yr | 3/4[b] | 7837 | Children + adults | TS | 2.3 | 31.5 | 16.7 | 22.9 | | 65.1 | 1/0[a] | Single | long-term trend, seasonality, temperature, humidity, day-of-week |
| 75, Neidell M, 2010 | Southern California, 8 yr | 4 | - | General | TS | | | | | | 175.7 | 1 | Cumulative | long-term trend, seasonality, temperature, humidity, day-of-week |
| 81, Fung KY, 2005 | London, Ontario, Canada, 5 yr | 4 | ? | Children + adults | TS | | 38.0 | | | | | 1 | Both | long-term trend, seasonality, temperature, humidity, day-of-week |
| 83, Hua J, 2014 | Shanghai, China, 7 yr | 4 | | Children | TS | | | 34.0 | | | | 1 | cumulative | long-term trend, seasonality, temperature, humidity |
| 9, Walters S, 1994 | Birmingham, UK, 2 yr | 2 | - | General | TS | | | | 39.1 | | | 1 | Cumulative | not declared |
| 35, Magas OK, 2007 | Oklahoma city metropolitan area, US, 3 yr | 1 | 1270 | Children | TS | | | | | - | | 0 | Single | temperature, day-of-week, humidity, holiday |
| 53, Barnett AG, 2005 | Australia - Brisbane, Canberra, Melbourne, Perth, Sydney, New Zealand - Auckland, Christchurch, 3 yr | 5 | - | Children | CC | 1.1 | 17.6 | 9.4 | 7.0 | 17.6 | 49.6 | 1 | Cumulative | long-term trend, seasonality, temperature, humidity, day for the week, holiday and influenza epidemics |

| Study | Location | Col4 | N | Population | Design |  |  |  |  |  |  |  | Lag | Adjustments |
|---|---|---|---|---|---|---|---|---|---|---|---|---|---|---|
| 17, Yang CY, 2007 | Taipei, 8 yr | 4 | 25602 | General | CC | 1.7 | 49.0 | | 12.3 | 63.0 | 44.0 | 1 | Cumulative | long-term trend, seasonality, temperature, humidity |
| 18, Tsai SS, 2006 | Kaohsiung, 8 yr | 4 | 17682 | General | CC | 1.00 | 76.7 | | 27.1 | 55.9 | 56.3 | 1 | Cumulative | long-term trends, seasonality, temperature, humidity |
| 37, Lin M, 2003 | Toronto, Ontario, Canada. 13 yr | 4 | 7319 | Children | CC | 1.5 | | | 15.3 | 51.8 | 65.1ᵉ | 1 | Both | long term trend, seasonality, temperature, humidity, day-of-week, |
| 44, Lin M, 2002 | Toronto, Canada, 14 yr | 4 | 7319 | Children | CC | | 30.2 | 18.0 | | | | 1 | Both | long-term trend, seasonality, temperature, humidity, day-of-week |
| 68, Iskandar A, 2012 | Copenhagen, Denmark, 8 yr | 4 | 8226 | Children | CC | | 26.2 | 10.3 | | 23.3 | | 1 | Cumulative | long-term trend, seasonality, temperature, humidity, day-of-week |
| 85, Cheng MH, 2014 | Taipei, China, 5 yr | 3 | 10,440 | General | CC | | | 30.0 | | | | 1 | Cumulative | long-term trend, seasonality, temperature, humidity |
| 86, Cai J, 2014 | Shanghai, China, 7y r | 4 | 15,678 | General | TS | | 88.0 | | 45.0 | 60.0 | | 1 | Both | long-term trend, seasonality, temperature, humidity, day-of-week |

| Study | Location/Period | Quality score | Sample size | Population | Type of study | Pollutants and concentration | | | | | | Measurement quality score (0-1 point) | Lag pattern (single day/ mean days) | Adjustments (long-term trend, seasonality, temperature, humidity, pressure, day for the week, holiday and influenza epidemics) |
|---|---|---|---|---|---|---|---|---|---|---|---|---|---|---|
| | | | | | | CO | PM$_{10}$ | PM$_{2.5}$ | SO$_2$ | NO$_2$ | O$_3$ | | | |
| **Studies with both emergency room visit and hospital admission indices** | | | | | | | | | | | | | | |
| 33, Chew FT, 1999 | Singapore, 5 yr | 4 | 6000(HA), 23000(ER) | Children | TS | | | | 38.1 | 18.9 | | 1 | Single | long-term trend, seasonality, temperature, humidity, day-of-week |
| 62, Sluaghter JC, 2005 | Spokane, Washington DC, US, 7 yr | 4 | 2191/2373 | General | TS | 1.6 - 3.8 | 7.9 - 41.9 | 4.2 - 20.2 | | | | 1 | Single | long-term trend, seasonality, temperature, humidity, day-of-week |
| 26, Smargiassi A, 2009 | 0.5–7.5 km of the refinery stacks, Montreal Canada, 8 yr | 4 | 1579/263 | Children | CC | | | | 12.3 | | | 1 | Both | long-term trends, seasonality, temperature, humidity, day-of-week |
| 63, Li S, 2001 | Detroit, Michigan, 2 yr | 4 | 7063 | Children | CC | 0.5 | | 15.0 | 10.8 | 34.4[c] | | 1 | Both | long-term trend, seasonality, temperature, humidity, day-of-week |

# Summary Statistics of Other Air Quality Components and Calculated p-values

(Note: Publ year=publication year; RR=relative risk; LCL=lower confidence limit; UCL=upper confidence limit; p-value calculated after Altman and Bland (2011); bold, italicized p-value <.05)

**$CO_2$**

| Study 1st author | Publ year | RR | LCL | UCL | *p-value* |
|---|---|---|---|---|---|
| Thompson AJ | 2001 | 1.025 | 1.000 | 1.055 | 0.0702351 |
| Fusco D | 2001 | 1.005 | 0.975 | 1.038 | 0.7676324 |
| Szyszkowicz M | 2008 | 1.076 | 1.032 | 1.120 | ***0.000482*** |
| Norris G | 1999 | 1.133 | 1.027 | 1.253 | ***0.013787*** |
| Hajat S | 1999 | 1.020 | 0.985 | 1.058 | 0.2811338 |
| Lee JT | 2006 | 1.186 | 0.894 | 1.532 | 0.2162465 |
| Lin M | 2003 | 1.080 | 1.000 | 1.176 | 0.0623584 |
| Lin M | 2003 | 1.000 | 0.888 | 1.096 | 1 |
| Krmpotic D | 2011 | 1.218 | 1.026 | 1.436 | ***0.021316*** |
| Lin M (ML) | 2004 | 1.096 | 0.984 | 1.226 | 0.1020115 |
| Lin M (MH) | 2004 | 1.096 | 0.968 | 1.226 | 0.128406 |
| Lin M (FL) | 2004 | 1.160 | 0.872 | 1.176 | 0.0513557 |
| Lin M (FH) | 2004 | 1.080 | 0.904 | 1.256 | 0.3649995 |
| Abe T (A) | 2007 | 1.163 | 0.960 | 1.409 | 0.122931 |
| Abe T (C) | 2007 | 1.019 | 0.953 | 1.089 | 0.5922572 |
| Pereira G (M) | 2010 | 2.215 | 1.243 | 3.465 | ***0.002413*** |
| Pereira G (F) | 2010 | 1.382 | 0.451 | 2.424 | 0.4596406 |
| Jalaludin BB | 2008 | 1.017 | 1.010 | 1.025 | ***0.0001*** |
| Lavigne E | 2012 | 1.000 | 0.720 | 1.280 | 1 |
| Stieb DM | 2009 | 0.997 | 0.974 | 1.021 | 0.8145069 |
| Sluaghter JC (E) | 2005 | 1.000 | 0.960 | 1.048 | 1 |
| Sluaghter JC (H) | 2005 | 1.016 | 0.936 | 1.104 | 0.7193821 |
| Li S | 2011 | 1.005 | 0.959 | 1.052 | 0.8436515 |
| Mar TF | 2010 | 1.000 | 0.966 | 1.034 | 1 |
| Sheppard L | 1999 | 1.052 | 1.026 | 1.078 | ***0.0001*** |
| Santus P | 2012 | 1.030 | 0.938 | 1.132 | 0.5488283 |
| Fletcher T | 2000 | 1.009 | 0.977 | 1.022 | 0.4438295 |
| Strickland MJ (W) | 2010 | 1.120 | 1.072 | 1.169 | ***0.0001*** |
| Strickland MJ (C) | 2010 | 1.004 | 0.975 | 1.033 | 0.7987727 |
| Son JY | 2013 | 1.003 | 0.957 | 1.048 | 0.905164 |
| Halonen JI | 2008 | 0.927 | 0.820 | 1.035 | 0.2034389 |
| Halonen JI | 2008 | 1.090 | 0.947 | 1.240 | 0.2118576 |
| Evans KA | 2013 | 1.420 | 0.037 | 3.395 | 0.7736967 |

**CO2 (continued)**

| Study 1st author | Publ year | RR | LCL | UCL | | *p-value* |
|---|---|---|---|---|---|---|
| Ito K | 2007 | 1.078 | 1.066 | 1.090 | | ***0.0001*** |
| Yang CY | 2007 | 1.417 | 1.270 | 1.578 | | ***0.0001*** |
| Yang CY | 2007 | 1.115 | 1.029 | 1.205 | | ***0.006901*** |
| Tsai SS | 2006 | 1.612 | 1.380 | 1.860 | | ***0.0001*** |
| Tsai SS | 2006 | 2.140 | 1.826 | 2.479 | | ***0.0001*** |
| Barnett AG | 2005 | 1.020 | 0.990 | 1.050 | | 0.188257 |
| Andersen ZJ | 2008 | 1.000 | 0.993 | 1.008 | | 1 |
| Lee JT | 2002 | 1.128 | 1.080 | 1.176 | | ***0.0001*** |
| Raun LH | 2014 | 1.155 | 1.000 | 1.258 | | ***0.013784*** |

**NO2**

| Study 1st author | Publ year | RR | LCL | UCL | p-value |
|---|---|---|---|---|---|
| Thompson AJ | 2001 | 1.039 | 1.015 | 1.063 | *0.001219* |
| Samoli E | 2010 | 1.011 | 0.990 | 1.029 | 0.2702846 |
| Fusco D | 2001 | 1.021 | 0.998 | 1.043 | 0.0643187 |
| Morgan G (C) | 1998 | 1.009 | 0.995 | 1.024 | 0.2235095 |
| Morgan G (A) | 1998 | 1.007 | 0.991 | 1.022 | 0.3812297 |
| Laurent O | 2008 | 1.025 | 0.990 | 1.062 | 0.1687421 |
| Anderson HR | 1998 | 1.006 | 1.002 | 1.010 | *0.003242* |
| Szyszkowicz M | 2008 | 1.020 | 1.008 | 1.032 | *0.001015* |
| Lee SL | 2006 | 1.016 | 1.009 | 1.023 | *0.0001* |
| Norris G | 1999 | 1.020 | 0.996 | 1.049 | 0.134508 |
| Hajat S | 1999 | 1.004 | 0.999 | 1.010 | 0.1534962 |
| Tenias JM | 1998 | 1.076 | 1.020 | 1.134 | *0.006746* |
| Chew FT | 1999 | 1.006 | 0.945 | 1.011 | 0.7413803 |
| Magas OK | 2007 | 1.133 | 1.047 | 1.227 | *0.002086* |
| Lee JT | 2006 | 1.018 | 1.006 | 1.033 | *0.008283* |
| Lin M (M) | 2003 | 1.018 | 0.996 | 1.044 | 0.1375644 |
| Lin M (F) | 2003 | 0.996 | 0.965 | 1.027 | 0.8126995 |
| Ko FWS | 2007 | 1.009 | 1.005 | 1.014 | *0.0001* |
| Krmpotic D | 2011 | 1.058 | 1.006 | 1.110 | *0.02447* |
| Lin M (ML) | 2004 | 1.097 | 1.030 | 1.172 | *0.004995* |
| Lin M (MH) | 2004 | 1.030 | 0.963 | 1.105 | 0.4068878 |
| Lin M (FL) | 2004 | 1.052 | 0.970 | 1.142 | 0.2255329 |
| Lin M (FH) | 2004 | 1.007 | 0.925 | 1.097 | 0.8819162 |
| Galan I | 2003 | 1.013 | 0.991 | 1.035 | 0.2464135 |
| Schouten JP | 1996 | 1.006 | 0.989 | 1.027 | 0.5450544 |
| Boutin-Forzano S | 2004 | 1.007 | 0.996 | 1.018 | 0.2124552 |
| Pereira G (M) | 2010 | 1.779 | 1.049 | 2.558 | *0.011266* |
| Pereira G (F) | 2010 | 1.195 | 0.562 | 1.925 | 0.582443 |
| Jalaludin BB | 2008 | 1.012 | 1.007 | 1.016 | *0.0001* |
| Jaffe DH | 2003 | 1.015 | 0.995 | 1.034 | 0.1291189 |
| Lavigne E | 2012 | 0.989 | 0.923 | 1.054 | 0.7568099 |
| Stieb DM | 2009 | 0.999 | 0.988 | 1.010 | 0.8685931 |
| Li S | 2011 | 1.019 | 1.003 | 1.036 | *0.022479* |
| Cirera L | 2012 | 1.026 | 1.004 | 1.049 | *0.021576* |
| Yamazaki S (W) | 2013 | 1.054 | 0.784 | 1.591 | 0.7833535 |
| Yamazaki S (C) | 2013 | 0.939 | 0.809 | 1.126 | 0.4645442 |
| Santus P | 2012 | 0.968 | 0.954 | 0.981 | *0.0001* |
| Fletcher T | 2000 | 1.003 | 0.998 | 1.009 | 0.287819 |
| Son JY | 2013 | 1.009 | 1.001 | 1.016 | *0.018085* |
| Halonen JI (A) | 2008 | 1.015 | 0.983 | 1.048 | 0.3681822 |
| Halonen JI (O) | 2008 | 1.026 | 0.983 | 1.073 | 0.2535308 |

**NO2 (continued)**

| Study 1st author | Publ year | RR | LCL | UCL | | p-value |
|---|---|---|---|---|---|---|
| Sunyer J (A) | 1997 | 1.006 | 1.001 | 1.011 | | *0.018198* |
| Sunyer J (C) | 1997 | 1.005 | 1.001 | 1.010 | | *0.028699* |
| Mehta AJ | 2012 | 1.019 | 0.849 | 1.223 | | 0.8504996 |
| Ito K | 2007 | 1.024 | 1.015 | 1.032 | | *0.0001* |
| Cassino C | 1999 | 0.990 | 0.951 | 1.029 | | 0.6298615 |
| Chardon B | 2007 | 0.997 | 0.967 | 1.027 | | 0.8553863 |
| Yang CY (W) | 2007 | 1.086 | 1.055 | 1.120 | | *0.0001* |
| Yang CY (C) | 2007 | 1.062 | 1.037 | 1.088 | | *0.0001* |
| Tsai SS (W) | 2006 | 1.074 | 1.032 | 1.122 | | *0.000859* |
| Tsai SS (C) | 2006 | 1.321 | 1.251 | 1.399 | | *0.0001* |
| Wong TW | 1999 | 1.026 | 1.010 | 1.042 | | *0.001305* |
| Petroeschevsky A | 2001 | 0.998 | 0.997 | 0.999 | | *0.000102* |
| Jazbec A (C) | 1999 | 1.137 | 1.005 | 1.287 | | *0.041512* |
| Jazbec A (A) | 1999 | 1.098 | 1.002 | 1.203 | | *0.044647* |
| Castellsague J (W) | 1995 | 1.018 | 1.004 | 1.032 | | *0.010979* |
| Castellsague J (C) | 1995 | 1.022 | 1.004 | 1.042 | | *0.021497* |
| Barnett AG | 2005 | 1.023 | 0.992 | 1.054 | | 0.1417651 |
| Barnett AG | 2005 | 1.055 | 1.016 | 1.096 | | *0.005655* |
| Medina S | 1997 | 1.041 | 1.026 | 1.053 | | *0.0001* |
| Iskandar A | 2012 | 1.075 | 1.030 | 1.119 | | *0.000662* |
| Strickland MJ | 2010 | 1.014 | 1.007 | 1.021 | | *0.0001* |
| Andersen ZJ | 2008 | 1.032 | 0.935 | 1.146 | | 0.5553011 |
| Lee JT | 2002 | 1.050 | 1.033 | 1.067 | | *0.0001* |
| Cai J | 2014 | 1.011 | 1.005 | 1.016 | | *0.0001* |
| Raun LH | 2014 | 1.073 | 1.042 | 1.104 | | *0.0001* |

**O3**

| Study 1st author | Publ year | RR | LCL | UCL | p-value |
|---|---|---|---|---|---|
| Thompson AJ | 2001 | 0.977 | 0.953 | 1.005 | 0.0856757 |
| Samoli E | 2010 | 0.969 | 0.934 | 1.005 | 0.091725 |
| Fusco D | 2001 | 1.016 | 0.987 | 1.046 | 0.2875781 |
| Anderson HR | 1998 | 1.003 | 1.000 | 1.005 | *0.018427* |
| Szyszkowicz M | 2008 | 1.028 | 1.013 | 1.043 | *0.000229* |
| Lee SL | 2006 | 1.010 | 1.002 | 1.019 | *0.020273* |
| Hajat S | 1999 | 0.997 | 0.990 | 1.004 | 0.4090144 |
| Tenias JM | 1998 | 1.063 | 1.014 | 1.114 | *0.010857* |
| Stieb DM | 1996 | 1.036 | 1.017 | 1.055 | *0.000175* |
| Lee JT | 2006 | 1.025 | 1.005 | 1.027 | *0.0001* |
| Galan I | 2003 | 1.039 | 1.010 | 1.068 | *0.007249* |
| Mohr LB (W) | 2008 | 0.981 | 0.953 | 1.014 | 0.2276172 |
| Mohr LB (C) | 2008 | 1.023 | 0.986 | 1.065 | 0.2501553 |
| Boutin-Forzano S | 2004 | 1.006 | 1.000 | 1.016 | 0.1398602 |
| Medina S | 1997 | 1.012 | 0.995 | 1.033 | 0.2139423 |
| Jaffe DH | 2003 | 1.014 | 1.000 | 1.028 | *0.048062* |
| Lavigne E | 2012 | 1.042 | 1.004 | 1.081 | *0.028827* |
| Stieb DM | 2009 | 1.000 | 0.990 | 1.011 | 1 |
| Wilson AM (P) | 2004 | 1.020 | 1.000 | 1.030 | *0.008634* |
| Wilson AM (M) | 2004 | 0.990 | 0.980 | 1.010 | 0.1926288 |
| Amancio CT | 2012 | 1.000 | 1.000 | 1.001 | 1 |
| Cirera L | 2012 | 0.971 | 0.938 | 1.006 | 0.099055 |
| Sheppard L | 1999 | 1.014 | 1.005 | 1.026 | *0.008407* |
| Yamazaki S (W) | 2013 | 1.054 | 0.970 | 1.154 | 0.2376168 |
| Yamazaki S (C) | 2013 | 0.956 | 0.860 | 1.080 | 0.4472306 |
| Babin S | 2008 | 1.004 | 1.001 | 1.008 | *0.024531* |
| Mar TF | 2009 | 1.047 | 1.005 | 1.089 | *0.024703* |
| Hernandez-Cadena L | 2000 | 1.004 | 0.997 | 1.012 | 0.2987379 |
| Fletcher T | 2000 | 1.000 | 0.991 | 1.011 | 1 |
| Morgan G (C) | 1998 | 0.997 | 0.992 | 1.002 | 0.242804 |
| Morgan G (A) | 1998 | 1.003 | 0.997 | 1.010 | 0.3709557 |
| Sunyer J (A) | 1997 | 1.003 | 0.987 | 1.040 | 0.8336614 |
| Sunyer J (C) | 1997 | 0.998 | 0.988 | 1.008 | 0.7085282 |
| Yamazaki S | 2009 | 1.009 | 1.005 | 1.110 | 0.7368412 |
| Chakraborty P | 2013 | 0.998 | 0.996 | 0.999 | *0.009064* |
| Evans KA | 2013 | 0.957 | 0.887 | 1.154 | 0.5232048 |
| Ito K | 2007 | 1.011 | 0.999 | 1.014 | *0.004056* |
| Yang CY (W) | 2007 | 1.012 | 0.986 | 1.039 | 0.3782693 |
| Yang CY (C) | 2007 | 1.073 | 1.046 | 1.104 | *0.0001* |
| Tsai SS (W) | 2006 | 1.067 | 1.046 | 1.089 | *0.0001* |
| Tsai SS (C) | 2006 | 1.047 | 1.017 | 1.081 | *0.003229* |

**O3 (continued)**

| Study 1st author | Publ year | RR | LCL | UCL | | *p-value* |
|---|---|---|---|---|---|---|
| Wong TW | 1999 | 1.031 | 1.017 | 1.460 | | 0.753609 |
| Laurent O | 2008 | 0.998 | 0.965 | 1.032 | | 0.914377 |
| Petroeschevsky A | 2001 | 1.004 | 1.002 | 1.007 | | ***0.00172*** |
| Lin M(M) | 2003 | 0.975 | 0.925 | 1.025 | | 0.3389074 |
| Lin M(F) | 2003 | 0.913 | 0.863 | 1.025 | | ***0.037762*** |
| Lin M(ML) | 2004 | 0.964 | 0.943 | 0.986 | | ***0.001316*** |
| Lin M(MH) | 2004 | 0.983 | 0.960 | 1.095 | | 0.6219986 |
| Lin M(FL) | 2004 | 1.026 | 0.993 | 1.066 | | 0.1566245 |
| Lin M(FH) | 2004 | 0.979 | 0.948 | 1.012 | | 0.2043825 |
| Ko FWS | 2007 | 1.010 | 1.006 | 1.014 | | ***0.0001*** |
| Schouten JP | 1996 | 1.009 | 0.993 | 1.027 | | 0.3009679 |
| Castellsague J (W) | 1995 | 0.996 | 0.976 | 1.018 | | 0.7223679 |
| Castellsague J (C) | 1995 | 1.022 | 0.999 | 1.046 | | 0.0631237 |
| Silverman RA | 2010 | 1.019 | 1.013 | 1.025 | | ***0.0001*** |
| Barnett AG | 2005 | 0.990 | 0.953 | 1.031 | | 0.6291404 |
| Barnett AG | 2005 | 0.980 | 0.940 | 1.025 | | 0.3663807 |
| Jalaludin BB | 2008 | 1.003 | 1.001 | 1.005 | | ***0.003287*** |
| Cassino C | 1999 | 1.087 | 1.033 | 1.146 | | ***0.001684*** |
| Neidell M | 2010 | 1.013 | 1.010 | 1.016 | | ***0.0001*** |
| Santus P | 2012 | 1.010 | 0.991 | 1.052 | | 0.5243647 |
| Yamazaki S | 2009 | 1.073 | 1.000 | 1.151 | | ***0.049155*** |
| Strickland MJ | 2010 | 1.010 | 1.005 | 1.015 | | ***0.000093*** |
| Son JY | 2013 | 1.004 | 0.999 | 1.009 | | 0.1160943 |
| Paulu C | 2008 | 1.009 | 1.000 | 1.023 | | 0.1224737 |
| Andersen ZJ | 2008 | 1.043 | 0.914 | 1.226 | | 0.5860823 |
| Evans KA | 2013 | 0.911 | 0.793 | 1.121 | | 0.2951183 |
| Lee JT | 2002 | 1.026 | 1.015 | 1.034 | | ***0.0001*** |
| Gleason JA | 2014 | 1.032 | 1.026 | 1.037 | | ***0.0001*** |
| Raun LH | 2014 | 1.009 | 1.002 | 1.016 | | ***0.011329*** |
| Sacks JD | 2014 | 1.004 | 1.000 | 1.009 | | 0.0803567 |

**PM10**

| Study 1st author | Publ year | RR | LCL | UCL | p-value |
|---|---|---|---|---|---|
| Thompson AJ | 2001 | 1.009 | 1.000 | 1.019 | 0.0616359 |
| Samoli E | 2010 | 1.025 | 1.001 | 1.051 | *0.046683* |
| Szyszkowicz M | 2008 | 1.018 | 1.001 | 1.036 | 0.0415255 |
| Lee SL | 2006 | 1.017 | 1.011 | 1.023 | *0.0001* |
| Norris G | 1999 | 1.121 | 1.043 | 1.198 | *0.001279* |
| Lee JT | 2006 | 1.016 | 1.008 | 1.023 | *0.0001* |
| Ko FWS | 2007 | 1.005 | 1.002 | 1.009 | *0.005019* |
| Krmpotic D | 2011 | 0.989 | 0.942 | 1.044 | 0.6863417 |
| Galan I | 2003 | 1.006 | 0.976 | 1.037 | 0.7120714 |
| Atkinson RW (C) | 2001 | 1.012 | 1.002 | 1.023 | *0.023976* |
| Atkinson RW (A) | 2001 | 1.011 | 1.003 | 1.018 | *0.003914* |
| Jalaludin BB | 2008 | 1.014 | 1.008 | 1.022 | *0.0001* |
| Jaffe DH | 2003 | 1.010 | 0.986 | 1.038 | 0.4566827 |
| Stieb DM | 2009 | 1.004 | 0.950 | 1.015 | 0.8246528 |
| Romero-Placeres M | 2004 | 0.998 | 0.943 | 1.003 | 0.9066903 |
| Chimonas MAR | 2007 | 1.006 | 1.001 | 1.013 | *0.048715* |
| Sluaghter JC (ER) | 2005 | 1.012 | 0.992 | 1.028 | 0.1908417 |
| Sluaghter JC (H) | 2005 | 1.012 | 0.980 | 1.048 | 0.4956615 |
| Amancio CT | 2012 | 1.006 | 0.999 | 1.013 | 0.0916967 |
| Sheppard L | 1999 | 1.025 | 1.010 | 1.040 | *0.000987* |
| Morgan G (C) | 2010 | 1.006 | 0.976 | 1.038 | 0.7165524 |
| Morgan G (A) | 2010 | 1.025 | 0.993 | 1.058 | 0.126933 |
| Yamazaki S (W) | 2013 | 0.968 | 0.818 | 1.146 | 0.7185001 |
| Yamazaki S (C) | 2013 | 1.022 | 0.902 | 1.158 | 0.7457867 |
| Santus P | 2012 | 1.005 | 0.987 | 1.024 | 0.6075507 |
| Babin S | 2008 | 0.990 | 0.960 | 1.010 | 0.4462646 |
| Hernandez-Cadena L | 2000 | 1.013 | 0.993 | 1.033 | 0.2012836 |
| Fletcher T | 2000 | 1.005 | 0.992 | 1.020 | 0.4922045 |
| Son JY | 2013 | 1.005 | 1.001 | 1.008 | *0.005062* |
| Delfino RJ | 1994 | 1.021 | 1.006 | 1.040 | *0.014173* |
| Ye F | 2001 | 1.003 | 1.001 | 1.004 | *0.0001* |
| Morgan G (C) | 1998 | 0.997 | 0.984 | 1.011 | 0.6765743 |
| Morgan G (A) | 1998 | 1.005 | 0.992 | 1.018 | 0.4586868 |
| Malig BJ | 2013 | 1.022 | 1.003 | 1.031 | *0.002* |
| Chakraborty P | 2013 | 1.000 | 1.000 | 1.003 | 1 |
| Fung KY | 2005 | 1.096 | 1.011 | 1.201 | *0.036618* |
| Chardon B | 2007 | 1.025 | 0.983 | 1.068 | 0.2457323 |
| Yang CY (W) | 2007 | 1.017 | 0.989 | 1.048 | 0.2570002 |
| Yang CY (C) | 2007 | 1.018 | 1.004 | 1.033 | *0.013982* |
| Tsai SS | 2006 | 1.048 | 1.025 | 1.075 | *0.000128* |
| Wong TW | 1999 | 1.015 | 1.002 | 1.028 | *0.022532* |

**PM10 (continued)**

| Study 1st author | Publ year | RR | LCL | UCL | | *p-value* |
|---|---|---|---|---|---|---|
| Laurent O | 2008 | 1.035 | 0.997 | 1.075 | | 0.073024 |
| Lin M | 2002 | 1.033 | 0.966 | 1.115 | | 0.3815059 |
| Barnett AG (0-4y) | 2005 | 1.023 | 1.007 | 1.039 | | *0.004424* |
| Barnett AG (4-14y) | 2005 | 1.025 | 1.001 | 1.051 | | *0.046683* |
| Iskandar A | 2012 | 1.052 | 1.022 | 1.090 | | *0.00209* |
| Andersen ZJ | 2008 | 1.015 | 0.946 | 1.092 | | 0.6974281 |
| Strickland MJ | 2010 | 1.014 | 1.002 | 1.026 | | *0.021146* |
| Lee JT | 2002 | 1.012 | 1.009 | 1.027 | | *0.008187* |
| Cadelis G | 2014 | 1.010 | 0.941 | 1.046 | | 0.7254741 |
| Cai J | 2014 | 1.001 | 0.998 | 1.005 | | 0.5870548 |

**SO2**

| Study 1st author | Publ year | RR | LCL | UCL | | p-value |
|---|---|---|---|---|---|---|
| Thompson AJ | 2001 | 1.007 | 1.003 | 1.011 | | ***0.000614*** |
| Wong TW | 1999 | 1.017 | 0.998 | 1.036 | | 0.0766384 |
| Fusco D | 2001 | 0.978 | 0.904 | 1.057 | | 0.5890392 |
| Anderson HR | 1998 | 1.016 | 1.005 | 1.028 | | 0.0059914 |
| Hajat S | 1999 | 1.018 | 1.001 | 1.034 | | 0.0308161 |
| Petroeschevsky A | 2001 | 0.998 | 0.996 | 1.000 | | ***0.049846*** |
| Ko FWS | 2007 | 1.004 | 0.998 | 1.011 | | 0.2287405 |
| Galan I | 2003 | 1.018 | 0.984 | 1.054 | | 0.3133554 |
| Schouten JP | 1996 | 0.980 | 0.970 | 0.992 | | ***0.000444*** |
| Medina S | 1997 | 1.026 | 1.010 | 1.041 | | ***0.000917*** |
| Lavigne E | 2012 | 1.044 | 0.933 | 1.133 | | 0.3916664 |
| Stieb DM | 2009 | 0.992 | 0.980 | 1.005 | | 0.2130706 |
| Romero-Placeres M | 2004 | 0.994 | 0.983 | 1.004 | | 0.2675815 |
| Wilson AM (P) | 2004 | 1.040 | 1.000 | 1.070 | | ***0.022883*** |
| Wilson AM (M) | 2004 | 1.020 | 0.990 | 1.060 | | 0.2587899 |
| Cirera L | 2012 | 1.060 | 1.014 | 1.110 | | ***0.011528*** |
| Romieu I | 1995 | 1.003 | 0.999 | 1.008 | | 0.1916958 |
| Santus P | 2012 | 0.950 | 0.808 | 1.104 | | 0.5301991 |
| Son JY | 2013 | 1.002 | 0.983 | 1.023 | | 0.8548452 |
| Sunyer J | 1997 | 0.999 | 0.992 | 1.007 | | 0.8059108 |
| Samoli E | 2010 | 1.060 | 1.009 | 1.113 | | ***0.019747*** |
| Lee SL | 2006 | 0.989 | 0.974 | 0.998 | | 0.0744936 |
| Norris G | 1999 | 1.006 | 0.985 | 1.026 | | 0.5770494 |
| Tenias JM | 1998 | 1.050 | 0.973 | 1.133 | | 0.2107013 |
| Stieb DM | 1996 | 0.997 | 0.992 | 1.006 | | 0.4080414 |
| Chew FT | 1999 | 1.015 | 1.007 | 1.023 | | ***0.000235*** |
| Lee JT | 2006 | 1.077 | 1.000 | 1.161 | | 0.0510455 |
| Lin M (M) | 2003 | 1.000 | 0.975 | 1.025 | | 1 |
| Lin M (F) | 2003 | 1.020 | 0.985 | 1.055 | | 0.2611838 |
| Lin M (ML) | 2004 | 1.021 | 0.934 | 1.106 | | 0.6426132 |
| Lin M (MH) | 2004 | 1.032 | 0.947 | 1.127 | | 0.4876276 |
| Lin M (FL) | 2004 | 1.053 | 0.947 | 1.170 | | 0.3438004 |
| Lin M (FH) | 2004 | 1.074 | 0.958 | 1.201 | | 0.2175988 |
| Mohr LB (W) | 2008 | 0.996 | 0.975 | 1.018 | | 0.7289469 |
| Mohr LB (C) | 2008 | 1.011 | 0.985 | 1.035 | | 0.3933581 |
| Abe T (A) | 2007 | 1.008 | 0.930 | 1.092 | | 0.8562393 |
| Abe T (C) | 2007 | 0.996 | 0.977 | 1.024 | | 0.751056 |
| Boutin-Forzano S | 2004 | 1.002 | 0.995 | 1.010 | | 0.6130637 |
| Jalaludin BB | 2008 | 1.070 | 1.031 | 1.105 | | ***0.000146*** |
| Jaffe DH | 2003 | 1.024 | 1.002 | 1.046 | | ***0.03026*** |
| Li S | 2011 | 1.027 | 1.007 | 1.048 | | ***0.008869*** |

**SO2 (continued)**

| Study 1st author | Publ year | RR | LCL | UCL | | p-value |
|---|---|---|---|---|---|---|
| Amancio CT | 2012 | 1.027 | 1.002 | 1.053 | | *0.03511* |
| Sheppard L | 1999 | 1.014 | 0.986 | 1.028 | | 0.1926567 |
| Fletcher T | 2000 | 1.039 | 0.993 | 1.091 | | 0.1109411 |
| Sunyer J (C) | 2003 | 1.013 | 1.004 | 1.022 | | *0.004425* |
| Sunyer J (A) | 2003 | 1.000 | 0.991 | 1.010 | | 1 |
| Smargiassi A (H) | 2009 | 1.169 | 1.000 | 1.360 | | *0.046151* |
| Smargiassi A (E) | 2009 | 1.048 | 0.976 | 1.121 | | 0.185688 |
| Evans KA | 2013 | 1.090 | 0.887 | 1.350 | | 0.4292189 |
| Ito K | 2007 | 1.033 | 1.030 | 1.043 | | *0.0001* |
| Cassino C | 1999 | 0.977 | 0.938 | 1.019 | | 0.2741459 |
| Yang CY (W) | 2007 | 1.091 | 1.001 | 1.187 | | *0.044792* |
| Yang CY (C) | 2007 | 0.978 | 0.920 | 1.038 | | 0.4793457 |
| Tsai SS (W) | 2006 | 1.011 | 0.973 | 1.050 | | 0.5853068 |
| Tsai SS (C) | 2006 | 1.113 | 1.044 | 1.190 | | *0.001395* |
| Laurent O | 2008 | 1.056 | 0.979 | 1.139 | | 0.1588227 |
| Castellsague J (W) | 1995 | 1.021 | 0.992 | 1.052 | | 0.1660837 |
| Castellsague J (C) | 1995 | 1.008 | 0.984 | 1.034 | | 0.5395251 |
| Barnett AG | 2005 | 1.017 | 1.004 | 1.031 | | *0.012717* |
| Barnett AG | 2005 | 1.013 | 0.964 | 1.084 | | 0.6791539 |
| Strickland MJ | 2010 | 1.004 | 0.998 | 1.009 | | 0.1539128 |
| Walters S (W) | 1994 | 1.012 | 0.987 | 1.448 | | 0.9105847 |
| Walters S (C) | 1994 | 1.037 | 1.000 | 1.076 | | 0.051474 |
| Cai J | 2014 | 1.013 | 1.005 | 1.021 | | *0.001398* |
| Raun LH | 2014 | 1.150 | 0.850 | 1.400 | | 0.275625 |

# Supplemental Information, SI 4

# Summary Statistics of Datasets from Meta-analysis of Selected Cancers in Petroleum Refinery Workers after Schnatter et al. (2018)

(Note: Base study=base study 1st author name in Schnatter et al. (2018); RR=relative risk; LCL=lower confidence limit; UCL=upper confidence limit; p-value calculated after Altman (2011); bold, italicized p-value <.05)

Chronic myeloid leukemia risk for petroleum refinery workers:

| Base study | RR | LCL | UCL | p-value |
|---|---|---|---|---|
| Collingwood 1996 | 0.53 | 0.07 | 3.74 | 0.54263 |
| Divine 1999a | 1.05 | 0.60 | 1.85 | 0.87478 |
| Gun 2006b | 1.09 | 0.45 | 2.61 | 0.85800 |
| Huebner 2004 | 1.68 | 0.88 | 3.23 | 0.11778 |
| Lewis 2000a | 1.08 | 0.35 | 3.35 | 0.90196 |
| Rushton 1993a | 0.89 | 0.50 | 1.61 | 0.70923 |
| Satin 1996 | 0.85 | 0.38 | 1.88 | 0.70346 |
| Satin 2002 | 0.45 | 0.14 | 1.39 | 0.17355 |
| Tsai 2007 | 0.66 | 0.21 | 2.05 | 0.48425 |
| Wong 2001a | 1.31 | 0.55 | 3.15 | 0.55549 |
| Wong 2001b | 1.96 | 0.49 | 7.84 | 0.34689 |
| Wongsrichanalia 1989 | 0.44 | 0.06 | 3.12 | 0.42318 |

Mesothelioma risk for petroleum refinery workers (based on mesothelioma subgroup analysis using Schnatter et al. (2018) 'Best Methods' dataset):

| Base study | RR | LCL | UCL | p-value |
|---|---|---|---|---|
| Devine 1999a | 2.97 | 2.21 | 3.99 | ***0.0001*** |
| Gamble 2000 | 2.43 | 1.35 | 4.39 | ***0.00321*** |
| Gun 2006a | 3.77 | 2.14 | 6.64 | ***0.0001*** |
| Honda 1995 | 2.00 | 1.04 | 3.84 | ***0.03720*** |
| Hornstra 1993 | 5.51 | 3.38 | 8.99 | ***0.0001*** |
| Huebner 2009 | 2.44 | 1.83 | 3.24 | ***0.0001*** |
| Kaplan 1986 | 2.41 | 1.26 | 4.64 | ***0.00817*** |
| Lewis 2000a | 8.68 | 5.77 | 13.06 | ***0.0001*** |
| Tsai 2003 | 2.16 | 0.70 | 6.69 | 0.18215 |
| Tsai 2007 | 2.50 | 1.63 | 3.83 | ***0.0001*** |

## Summary Statistics of Datasets from Meta-analysis of Elderly Long-term Exercise Training−Mortality & Morbidity Risk after de Souto Barreto et al. (2019)

(Note: Base study=base study 1st author name in de Souto Barreto et al. (2019); RR=relative risk; LCL=lower confidence limit; UCL=upper confidence limit; p-value calculated after Altman (2011); bold, italicized p-value <.05)

| Outcome | No. | Base Study ID (n=69) | RR | LCL | UCL | *p-value* |
|---|---|---|---|---|---|---|
| Mortality | 1 | Belardinelli et al. 2012 | 0.38 | 0.13 | 1.15 | 0.08153 |
|  | 2 | Barnett et al. 2003 | 0.14 | 0.01 | 2.63 | 0.16736 |
|  | 3 | O'Connor et al. 2009 | 0.96 | 0.80 | 1.16 | 0.67980 |
|  | 4 | Campbell et al. 1997 | 0.50 | 0.09 | 2.70 | 0.43245 |
|  | 5 | El−Khoury et al. 2015 | 0.84 | 0.26 | 2.72 | 0.78350 |
|  | 6 | Galvão et al. 2014 | 3.00 | 0.13 | 71.92 | 0.50544 |
|  | 7 | Gianoudis et al. 2014 | 1.00 | 0.06 | 15.72 | 1.00000 |
|  | 8 | Hewitt et al. 2018 | 1.02 | 0.52 | 2.03 | 0.95866 |
|  | 9 | Karinkanta et al. 2007 | 0.33 | 0.01 | 7.93 | 0.52568 |
|  | 10 | Kemmler et al. 2010 | 0.33 | 0.01 | 8.10 | 0.52704 |
|  | 11 | King et al. 2002 | 0.32 | 0.01 | 7.68 | 0.51168 |
|  | 12 | Kovács et al. 2013 | 0.40 | 0.14 | 1.17 | 0.09039 |
|  | 13 | Lam et al. 2012 | 0.64 | 0.06 | 7.05 | 0.72673 |
|  | 14 | Lam et al. 2015 | 0.30 | 0.03 | 2.82 | 0.30309 |
|  | 15 | Lord et al. 2003 | 4.84 | 0.55 | 42.33 | 0.15519 |
|  | 16 | Merom et al. 2015 | 1.36 | 0.22 | 8.23 | 0.75225 |
|  | 17 | Pahor et al. 2006 | 0.99 | 0.14 | 6.97 | 0.99276 |
|  | 18 | Pahor et al. 2014/Gill et al. 2016 | 1.14 | 0.76 | 1.71 | 0.53739 |
|  | 19 | Patil et al. 2015 | 0.11 | 0.01 | 2.04 | 0.10355 |
|  | 20 | Pitkälä et al. 2013 | 0.25 | 0.06 | 1.14 | 0.06455 |
|  | 21 | Prescott et al. 2008 | 0.42 | 0.08 | 2.12 | 0.30363 |
|  | 22 | Rejeski et al. 2017 | 0.34 | 0.01 | 8.16 | 0.53914 |
|  | 23 | Rolland et al. 2007 | 0.88 | 0.34 | 2.28 | 0.80441 |
|  | 24 | Sherrington et al. 2014 | 1.10 | 0.46 | 2.63 | 0.84135 |
|  | 25 | Underwood et al. 2013 | 1.06 | 0.84 | 1.35 | 0.64301 |
|  | 26 | Van Uffelen et al. 2008 | 0.36 | 0.01 | 8.72 | 0.56573 |
|  | 27 | von Stengel et al. 2011 | 0.34 | 0.01 | 8.15 | 0.53907 |
|  | 28 | Voukelatos et al. 2015 | 9.09 | 0.49 | 167.75 | 0.13843 |
|  | 29 | Wolf et al. 2003 | 0.97 | 0.14 | 6.86 | 0.97786 |

| Outcome | No. | Base Study ID (n=69) | RR | LCL | UCL | *p-value* |
|---|---|---|---|---|---|---|
| Hospitalization | 30 | Belardinelli et al. 2012 | 0.30 | 0.15 | 0.62 | ***0.00092*** |
| | 31 | O'Connor et al. 2009 | 0.97 | 0.91 | 1.03 | 0.34039 |
| | 32 | Hambrecht et al. 2004 | 0.16 | 0.02 | 1.31 | 0.08551 |
| | 33 | Hewitt et al. 2018 | 0.64 | 0.27 | 1.50 | 0.31209 |
| | 34 | Kovács et al. 2013 | 2.00 | 0.19 | 21.21 | 0.57619 |
| | 35 | Messier et al. 2013 | 8.54 | 0.46 | 157.06 | 0.14992 |
| | 36 | Mustata et al. 2011 | 0.33 | 0.02 | 7.32 | 0.47075 |
| | 37 | Pahor et al. 2006 | 0.99 | 0.68 | 1.44 | 0.96195 |
| | 38 | Pahor et al. 2014/Gill et al. 2016 | 1.10 | 0.99 | 1.22 | 0.07332 |
| | 39 | Pitkala et al. 2013 | 0.78 | 0.55 | 1.12 | 0.17166 |
| | 40 | Rejeski et al. 2017 | 3.04 | 0.13 | 73.46 | 0.50161 |
| | 41 | Rolland et al. 2007 | 1.82 | 0.95 | 3.49 | 0.07083 |
| Injurious falls | 42 | Barnett et al. 2003 | 0.77 | 0.48 | 1.21 | 0.27108 |
| | 43 | Campbell et al. 1997 | 0.67 | 0.45 | 1.00 | ***0.04892*** |
| | 44 | El−Khoury et al. 2015 | 0.90 | 0.78 | 1.05 | 0.16541 |
| | 45 | Hewitt et al. 2018 | 0.58 | 0.42 | 0.81 | ***0.00120*** |
| | 46 | MacRae et al. 1994 | 0.16 | 0.01 | 2.92 | 0.20731 |
| | 47 | Pahor et al. 2014/Gill et al. 2016 | 0.89 | 0.66 | 1.20 | 0.45350 |
| | 48 | Patil et al. 2015 | 0.51 | 0.31 | 0.84 | ***0.00810*** |
| | 49 | Pitkälä et al. 2013 | 0.65 | 0.39 | 1.09 | 0.10016 |
| | 50 | Reinsch et al. 1992 | 1.46 | 0.37 | 5.81 | 0.60232 |
| Fractures | 51 | Belardinelli et al. 2012 | 0.19 | 0.01 | 3.89 | 0.27847 |
| | 52 | O'Connor et al. 2009 | 0.60 | 0.32 | 1.11 | 0.10725 |
| | 53 | El−Khoury et al. 2015 | 0.88 | 0.60 | 1.25 | 0.50488 |
| | 54 | Gianoudis et al. 2014 | 3.00 | 0.12 | 72.57 | 0.51161 |
| | 55 | Hewitt et al. 2018 | 0.80 | 0.20 | 3.11 | 0.76275 |
| | 56 | Karinkanta et al. 2007 | 1.00 | 0.15 | 6.73 | 1.00000 |
| | 57 | Kemmler et al. 2010 | 0.49 | 0.19 | 1.25 | 0.13795 |
| | 58 | Kovács et al. 2013 | 3.00 | 0.13 | 71.56 | 0.50509 |
| | 59 | Lam et al. 2012 | 1.27 | 0.06 | 28.95 | 0.88844 |
| | 60 | Pahor et al. 2014/Gil et al. 2016 | 0.87 | 0.63 | 1.19 | 0.39774 |
| | 61 | Patil et al. 2015 | 0.66 | 0.28 | 1.59 | 0.35403 |
| | 62 | Pitkälä et al. 2013 | 1.00 | 0.26 | 3.84 | 1.00000 |
| | 63 | Reinsch et al. 1992 | 0.45 | 0.04 | 4.78 | 0.52344 |
| | 64 | Rolland et al. 2007 | 2.50 | 0.50 | 12.44 | 0.26692 |
| | 65 | Sherrington et al. 2014 | 0.92 | 0.46 | 1.85 | 0.82585 |
| | 66 | Underwood et al. 2013 | 1.05 | 0.63 | 1.74 | 0.86094 |
| | 67 | Villareal et al. 2011 | 0.52 | 0.05 | 5.39 | 0.59601 |
| | 68 | von Stengel et al. 2011 | 0.58 | 0.18 | 1.87 | 0.36778 |
| | 69 | Wolf et al. 2003 | 0.78 | 0.17 | 3.67 | 0.76405 |

# Supplemental Information, SI 6

# Summary Statistics of Datasets from Meta-analysis of Smoking−Squamous Cell Carcinoma Risk after Lee et al. (2012)

(Note: Base study=base study 1st author name in Lee et al. (2012); RR=relative risk; LCL=lower confidence limit; UCL=upper confidence limit; p-value calculated after Altman (2011); bold, italicized p-value <.05)

| Place | No. | Base Study ID (n=102) | RR | LCL | UCL | *p-value* |
|---|---|---|---|---|---|---|
| USA | 1 | 1948 WYNDE4 m | 12.79 | 6.19 | 26.14 | ***0.0001*** |
| | 2 | 1948 WYNDE4 f | 2.82 | 2.55 | 13.31 | ***0.01380*** |
| | 3 | 1949 BRESLO c | 3.69 | 2.06 | 6.62 | ***0.0001*** |
| | 4 | 1952 HAMMON m | 16.88 | 6.29 | 45.29 | ***0.0001*** |
| | 5 | 1955 HAENSZ f | 3.00 | 1.90 | 4.73 | ***0.0001*** |
| | 6 | 1957 BYERS1 m | 8.29 | 5.29 | 13.00 | ***0.0001*** |
| | 7 | 1960 LOMBA2 f | 4.24 | 2.40 | 7.50 | ***0.0001*** |
| | 8 | 1962 WYNDE2 m | 19.72 | 6.21 | 62.59 | ***0.0001*** |
| | 9 | 1964 OSANN2 f | 35.10 | 4.80 | 256.00 | ***0.00048*** |
| | 10 | 1966 WYNDE3 m | 18.29 | 5.71 | 58.56 | ***0.0001*** |
| | 11 | 1966 WYNDE3 f | 6.79 | 2.45 | 18.82 | ***0.00025*** |
| | 12 | 1968 HINDS f | 16.13 | 7.66 | 33.97 | ***0.0001*** |
| | 13 | 1969 STAYNE m | 3.47 | 2.17 | 5.56 | ***0.0001*** |
| | 14 | 1969 WYNDE6 m | 18.59 | 12.74 | 27.13 | ***0.0001*** |
| | 15 | 1969 WYNDE6 f | 32.37 | 17.66 | 59.35 | ***0.0001*** |
| | 16 | 1975 COMSTO m | 8.07 | 1.91 | 34.02 | ***0.00452*** |
| | 17 | 1975 COMSTO f | 46.20 | 2.74 | 778.83 | ***0.00784*** |
| | 18 | 1976 BUFFLE m | 14.03 | 4.73 | 41.61 | ***0.0001*** |
| | 19 | 1976 BUFFLE f | 13.04 | 3.99 | 42.66 | ***0.0001*** |
| | 20 | 1979 CORREA c | 28.30 | 18.60 | 43.20 | ***0.0001*** |
| | 21 | 1979 SIEMIA m | 22.70 | 6.90 | 75.20 | ***0.0001*** |
| | 22 | 1980 DORGAN m | 18.90 | 7.00 | 51.30 | ***0.0001*** |
| | 23 | 1980 DORGAN f | 11.10 | 7.20 | 17.10 | ***0.0001*** |
| | 24 | 1981 JAIN m | 18.00 | 5.50 | 111.00 | ***0.00018*** |
| | 25 | 1981 JAIN f | 25.50 | 7.93 | 156.00 | ***0.0001*** |
| | 26 | 1981 WU f | 24.29 | 3.40 | 173.76 | ***0.00153*** |
| | 27 | 1983 BAND m | 37.45 | 17.60 | 79.58 | ***0.0001*** |
| | 28 | 1984 BROWN2 m | 11.10 | 9.50 | 12.90 | ***0.0001*** |
| | 29 | 1984 BROWN2 f | 20.10 | 16.40 | 24.80 | ***0.0001*** |

**(continued)**

| Place | No. | Base Study ID (n=102) | RR | LCL | UCL | *p-value* |
|---|---|---|---|---|---|---|
| USA | 30 | 1984 OSANN m | 36.10 | 17.80 | 73.30 | ***0.0001*** |
|  | 31 | 1984 OSANN f | 26.40 | 14.50 | 48.10 | ***0.0001*** |
|  | 32 | 1984 SCHWAR m1 | 32.81 | 4.48 | 240.23 | ***0.0001*** |
|  | 33 | 1984 SCHWAR m2 | 1.81 | 0.50 | 6.78 | 0.37881 |
|  | 34 | 1984 SCHWAR f1 | 43.23 | 2.60 | 718.15 | ***0.00862*** |
|  | 35 | 1984 SCHWAR f2 | 62.61 | 3.64 | 1076.10 | ***0.00441*** |
|  | 36 | 1985 KHUDER m | 7.82 | 3.87 | 15.77 | ***0.0001*** |
|  | 37 | 1986 ANDERS f | 25.57 | 10.29 | 63.56 | ***0.0001*** |
|  | 38 | 1989 HEGMAN c | 30.80 | 12.48 | 76.03 | ***0.0001*** |
| Europe | 39 | 1947 ORMOS m | 10.14 | 2.41 | 42.79 | ***0.00165*** |
|  | 40 | 1948 DOLL m | 13.17 | 4.12 | 42.10 | ***0.0001*** |
|  | 41 | 1948 DOLL f | 2.13 | 1.06 | 4.27 | ***0.03311*** |
|  | 42 | 1948 KREYBE m | 10.87 | 3.47 | 34.04 | ***0.0001*** |
|  | 43 | 1948 KREYBE f | 2.29 | 0.89 | 5.88 | 0.08506 |
|  | 44 | 1954 STASZE m | 57.77 | 3.58 | 933.17 | ***0.00430*** |
|  | 45 | 1954 STASZE f | 32.45 | 1.32 | 800.04 | ***0.03297*** |
|  | 46 | 1959 TIZZAN c | 2.70 | 1.99 | 3.67 | ***0.0001*** |
|  | 47 | 1964 ENGELA m | 6.45 | 1.97 | 21.11 | ***0.00211*** |
|  | 48 | 1966 TOKARS c | 6.80 | 1.20 | 38.70 | ***0.03026*** |
|  | 49 | 1971 NOU m | 27.17 | 6.60 | 11.85 | ***0.0001*** |
|  | 50 | 1971 NOU f | 7.09 | 1.35 | 37.19 | ***0.02043*** |
|  | 51 | 1972 DAMBER m | 11.80 | 6.40 | 23.00 | ***0.0001*** |
|  | 52 | 1975 ABRAHA m | 92.66 | 5.77 | 1488.21 | ***0.00143*** |
|  | 53 | 1975 ABRAHA f | 5.35 | 2.22 | 12.90 | ***0.00021*** |
|  | 54 | 1976 LUBIN2 m | 16.66 | 12.69 | 21.86 | ***0.0001*** |
|  | 55 | 1976 LUBIN2 f | 5.78 | 4.34 | 7.71 | ***0.0001*** |
|  | 56 | 1977 ALDERS m | 14.70 | 3.40 | 63.64 | ***0.00035*** |
|  | 57 | 1977 ALDERS f | 6.09 | 2.68 | 13.82 | ***0.0001*** |
|  | 58 | 1979 BARBON m | 14.52 | 6.35 | 33.20 | ***0.0001*** |
|  | 59 | 1979 DOSEME m | 3.60 | 2.60 | 5.00 | ***0.0001*** |
|  | 60 | 1980 JEDRYC m | 12.84 | 5.58 | 29.55 | ***0.0001*** |
|  | 61 | 1983 SVENSS f | 12.62 | 3.97 | 40.14 | ***0.0001*** |
|  | 62 | 1985 BECHER f | 10.69 | 2.43 | 47.00 | ***0.00177*** |
|  | 63 | 1987 KATSOU f | 6.11 | 2.69 | 13.87 | ***0.0001*** |
|  | 64 | 1988 JAHN m | 23.03 | 7.29 | 72.81 | ***0.0001*** |
| Asia | 65 | 1961 ISHIMA c | 21.00 | 3.38 | 868.40 | ***0.03122*** |
|  | 66 | 1964 JUSSAW m | 25.43 | 13.87 | 46.63 | ***0.0001*** |
|  | 67 | 1965 MATSUD m | 39.01 | 5.44 | 279.84 | ***0.00029*** |
|  | 68 | 1976 CHAN m | 15.22 | 3.61 | 64.12 | ***0.00023*** |
|  | 69 | 1976 CHAN f | 6.44 | 3.44 | 12.06 | ***0.0001*** |
|  | 70 | 1976 LAMWK2 m | 6.89 | 2.65 | 17.90 | ***0.0001*** |
|  | 71 | 1976 LAMWK2 f | 6.49 | 3.27 | 12.88 | ***0.0001*** |

**(continued)**

| Place | No. | Base Study ID (n=102) | RR | LCL | UCL | *p-value* |
|---|---|---|---|---|---|---|
| Asia | 72 | 1976 TSUGAN m | 14.55 | 0.75 | 283.37 | 0.07657 |
|  | 73 | 1978 ZHOU m | 3.14 | 1.90 | 5.18 | ***0.0001*** |
|  | 74 | 1978 ZHOU f | 3.81 | 1.50 | 9.68 | ***0.00496*** |
|  | 75 | 1981 KOO f | 4.15 | 2.46 | 6.98 | ***0.0001*** |
|  | 76 | 1981 LAMWK f | 10.54 | 4.19 | 26.52 | ***0.0001*** |
|  | 77 | 1981 XU3 m | 5.90 | 1.69 | 20.57 | ***0.00540*** |
|  | 78 | 1981 XU3 f | 25.67 | 4.99 | 131.94 | ***0.00012*** |
|  | 79 | 1982 ZHENG m | 16.82 | 6.05 | 46.71 | ***0.0001*** |
|  | 80 | 1982 ZHENG f | 5.45 | 3.11 | 9.54 | ***0.0001*** |
|  | 81 | 1983 LAMTH f | 8.10 | 4.16 | 15.77 | ***0.0001*** |
|  | 82 | 1984 GAO m | 8.40 | 4.70 | 15.00 | ***0.0001*** |
|  | 83 | 1984 GAO f | 7.20 | 4.60 | 11.10 | ***0.0001*** |
|  | 84 | 1984 LUBIN m | 6.33 | 2.29 | 17.45 | ***0.00040*** |
|  | 85 | 1985 CHOI m | 5.45 | 2.34 | 12.67 | ***0.0001*** |
|  | 86 | 1985 CHOI f | 6.94 | 2.68 | 17.96 | ***0.0001*** |
|  | 87 | 1985 WUWILL f | 4.20 | 3.00 | 5.90 | ***0.0001*** |
|  | 88 | 1986 SOBUE m | 17.88 | 7.82 | 40.87 | ***0.0001*** |
|  | 89 | 1986 SOBUE f | 8.74 | 5.09 | 15.02 | ***0.0001*** |
|  | 90 | 1988 WAKAI m | 8.61 | 2.08 | 35.72 | ***0.00305*** |
|  | 91 | 1988 WAKAI f | 25.23 | 6.87 | 92.66 | ***0.0001*** |
|  | 92 | 1990 FAN c | 11.68 | 5.04 | 27.04 | ***0.0001*** |
|  | 93 | 1990 GER c | 3.19 | 1.08 | 9.42 | ***0.03547*** |
|  | 94 | 1990 LUO c | 10.90 | 2.50 | 47.90 | ***0.00157*** |
|  | 95 | 1991 KIHARA c | 26.97 | 10.84 | 67.08 | ***0.0001*** |
|  | 96 | 1997 SEOW f | 17.50 | 6.95 | 44.09 | ***0.0001*** |
| Other | 97 | 1978 JOLY m | 31.21 | 7.69 | 126.68 | ***0.0001*** |
|  | 98 | 1978 JOLY f | 18.56 | 7.74 | 44.51 | ***0.0001*** |
|  | 99 | 1987 PEZZOT m | 62.74 | 3.86 | 1019.50 | ***0.00367*** |
|  | 100 | 1991 SUZUK2 c | 31.00 | 4.20 | 227.00 | ***0.00078*** |
|  | 101 | 1993 DESTE2 m | 13.20 | 4.70 | 37.10 | ***0.0001*** |
|  | 102 | 1994 MATOS m | 8.08 | 2.58 | 25.50 | ***0.00038*** |